\documentclass[aps,prb,reprint,superscriptaddress,showpacs,showkeys,preprintnumbers]{revtex4-1}

\usepackage[utf8]{inputenc}
\usepackage{graphicx}
\usepackage{bm}
\usepackage{amsmath}
\usepackage{amssymb}
\usepackage{hyperref}
\usepackage{subfigure}
\usepackage{float}
\usepackage{multirow}
\usepackage{booktabs}
\usepackage{varioref}
\usepackage{xr-hyper}
\usepackage[dvipsnames]{xcolor}
\usepackage{nicefrac}
\usepackage{xfrac}
\usepackage{hyperref}
\hypersetup{colorlinks,linkcolor=blue,urlcolor=blue,citecolor=blue}
\usepackage{ulem}
\usepackage{siunitx}
\usepackage{graphicx}
\usepackage{dcolumn}
\usepackage{braket}
\usepackage{wasysym}
\usepackage{textcomp}
\usepackage{siunitx}

\newcommand{\LNOftt}{La$_4$Ni$_3$O$_{10}$}

\newcommand{\LNOfts}{La$_3$Ni$_2$O$_{7}$}

\begin{document}

\author{Ying~Chan}
\affiliation{Department of Physics, The Chinese University of Hong Kong, Shatin, Hong Kong, China}
\affiliation{State Key Laboratory of  Quantum Information Technologies and Materials, The Chinese University of Hong Kong, Shatin, Hong Kong, China}

\author{Yuehong~Li}
\affiliation{Department of Physics, The Chinese University of Hong Kong, Shatin, Hong Kong, China}
\affiliation{State Key Laboratory of  Quantum Information Technologies and Materials, The Chinese University of Hong Kong, Shatin, Hong Kong, China}

\author{Yujie~Yan}
\affiliation{Department of Physics, The Chinese University of Hong Kong, Shatin, Hong Kong, China}
\affiliation{State Key Laboratory of  Quantum Information Technologies and Materials, The Chinese University of Hong Kong, Shatin, Hong Kong, China}

\author{Xunyang~Hong}
\affiliation{Department of Physics, The Chinese University of Hong Kong, Shatin, Hong Kong, China}
\affiliation{Physik-Institut, Universit\"{a}t Z\"{u}rich, Winterthurerstrasse 190, CH-8057 Z\"{u}rich, Switzerland}

\author{Tianren~Wang}
\affiliation{Department of Physics, The Chinese University of Hong Kong, Shatin, Hong Kong, China}
\affiliation{State Key Laboratory of  Quantum Information Technologies and Materials, The Chinese University of Hong Kong, Shatin, Hong Kong, China}

\author{Marli dos Reis Cantarino}
\affiliation{European Synchrotron Radiation Facility, BP 220, F-38043 Grenoble Cedex, France}

\author{Yinghao Zhu}
\affiliation{State Key Laboratory of Surface Physics and Department of Physics, Fudan University, Shanghai,
200433, China}

\author{Enkang Zhang}
\affiliation{State Key Laboratory of Surface Physics and Department of Physics, Fudan University, Shanghai,
200433, China}

\author{Lixing Chen}
\affiliation{State Key Laboratory of Surface Physics and Department of Physics, Fudan University, Shanghai,
200433, China}

\author{Jun Okamoto}
\affiliation{National Synchrotron Radiation Research Center, Hsinchu 30076, Taiwan}

\author{Hsiao-Yu Huang}
\affiliation{National Synchrotron Radiation Research Center, Hsinchu 30076, Taiwan}

\author{Di-Jing Huang}
\affiliation{National Synchrotron Radiation Research Center, Hsinchu 30076, Taiwan}

\author{N.~B.~Brookes}
\affiliation{European Synchrotron Radiation Facility, 71 Avenue des Martyrs, 38043 Grenoble, France}

\author{Johan~Chang}
\affiliation{Physik-Institut, Universit\"{a}t Z\"{u}rich, Winterthurerstrasse 190, CH-8057 Z\"{u}rich, Switzerland}

\author{Yao Shen}
\email{yshen@iphy.ac.cn}
\affiliation{Beijing National Laboratory for Condensed Matter Physics, Institute of Physics, Chinese Academy of Sciences, Beijing 100190, China}
\affiliation{School of Physical Sciences, University of Chinese Academy of Sciences, Beijing 100049, China}

\author{Jun~Zhao}
\email{zhaoj@fudan.edu.cn}
\affiliation{State Key Laboratory of Surface Physics and Department of Physics, Fudan University, Shanghai,
200433, China}

\author{Qisi~Wang}
\email{qwang@cuhk.edu.hk}
\affiliation{Department of Physics, The Chinese University of Hong Kong, Shatin, Hong Kong, China}
\affiliation{State Key Laboratory of  Quantum Information Technologies and Materials, The Chinese University of Hong Kong, Shatin, Hong Kong, China}

\title{Collective spin excitations in trilayer nickelate La$_4$Ni$_3$O$_{10}$}

\begin{abstract}
Ruddlesden-Popper (RP) nickelates have recently emerged as a new family of high-temperature superconductors.
In bilayer RP nickelates, magnetic excitations with large exchange couplings have been observed, supporting a spin-mediated pairing mechanism.
Whether comparable spin correlations persist in trilayer nickelates, however, remains unknown. Here, we present a Ni $L$-edge resonant inelastic X-ray scattering (RIXS) study of La$_4$Ni$_3$O$_{10}$ single crystals. 
While the orbital excitations remain similar to those of La$_3$Ni$_2$O$_{7}$, 
the collective spin excitations in La$_4$Ni$_3$O$_{10}$ exhibit a comparable bandwidth of about $60$ meV but substantially suppressed spectral weight, implying a weaker electronic correlation in the trilayer compounds. Our results underscore the three-dimensional and multi-orbital electronic character in La$_4$Ni$_3$O$_{10}$, highlighting important differences from the bilayer nickelates. These findings provide crucial insights into the evolution of magnetism across the RP nickelate family and its connection to superconductivity.
\end{abstract}

\maketitle

Understanding the electronic and magnetic interactions is central to the study of unconventional superconductivity, as these interactions may mediate electron pairing.
In cuprates, the low-energy physics is largely captured by a single‐band Hubbard model derived primarily from the Cu $d_{x^2-y^2}$ orbital hybridized with O $2p$ states, with the essential electronic structure residing in the quasi-two-dimensional CuO$_2$ planes~\cite{lee_doping_2006}. 
Within this framework, superconducting pairing is widely believed to arise from strong electronic correlations and antiferromagnetic spin fluctuations inherited from the Mott-insulating parent compounds~\cite{lee_doping_2006,keimer_from_2015}.
However, whether unconventional high-temperature superconductivity can be realized beyond the cuprate paradigm remains an open question.

The recent discovery of high-temperature superconductivity in Ruddlesden-Popper (RP) nickelates provides new insights into this question~\cite{sun_signatures_2023,zhu_superconductivity_2024}.
Unlike cuprates, the RP phase nickelates $RE_{n+1}$Ni$_n$O$_{3n+1}$ ($RE=$ rare earth), consist of $n$-layers of NiO$_6$ octahedra stacked along the $c$-axis, and exhibit enhanced three-dimensional character (Fig.~\href{fig:fig1}{1\textbf{a}}). 
Consequently, multi-band electronic structures are observed for both
bilayer ($n=2$) and trilayer ($n=3$) nickelates with $d_{x^2-y^2}$ and $d_{z^2}$ orbitals at or near the Fermi level, as revealed by angle-resolved photoemission spectroscopy (ARPES) measurements~\cite{li_fermiology_2017, yang_orbital_2024, au-yeung_universal_2025, li_angle-resolved_2025}.
Electron hopping between $d_{z^2}$ orbitals across apical oxygens potentially mediates a strong interlayer exchange ($J_z$) in La$_3$Ni$_2$O$_{7}$, which has been suggested by recent resonant inelastic X-ray scattering (RIXS) and neutron scattering studies~\cite{chen_electronic_2024, zhong_spin_2025, xie_strong_2024}, revealing collective spin excitations topping around 70~meV.
The active involvement of the $d_{z^2}$ orbital also has profound influence on superconductivity, as suggested by different theoretical proposals~\cite{liu_swave_2023, yang_possible_2023, zhang_trends_2023,lechermann_electronic_2023,heier_competing_2024,jiang_high_2024}, although its detailed role is still under debate.

Despite the similar structural and orbital motifs, trilayer and bilayer nickelates display significant distinctions. While the nominal valence of Ni ions in both compounds lies between 3$d^7$ and 3$d^8$, the additional NiO$_{6}$ layer in trilayer nickelates introduces a different local crystal-field environment, which inevitably modifies the spin states and magnetic interactions.
In fact, an incommensurate spin order has been identified in La$_4$Ni$_3$O$_{10}$~\cite{zhang_intertwined_2020}, different from the spin-stripe order found in La$_3$Ni$_2$O$_{7}$~\cite{chen_electronic_2024,ren_resolving_2025,gupta_anisotropic_2025}.
Meanwhile, the maximum superconducting transition temperature ($T_c$) achieved in bulk La$_3$Ni$_2$O$_{7}$ under pressure (80~K) significantly exceeds that of La$_4$Ni$_3$O$_{10}$ (30~K).
The similarities and differences between bilayer and trilayer nickelates provide a comparative platform for identifying the key ingredients governing nickelate superconductivity.

\begin{figure*}[t!]
    \centering
 \includegraphics[width=.99\textwidth]{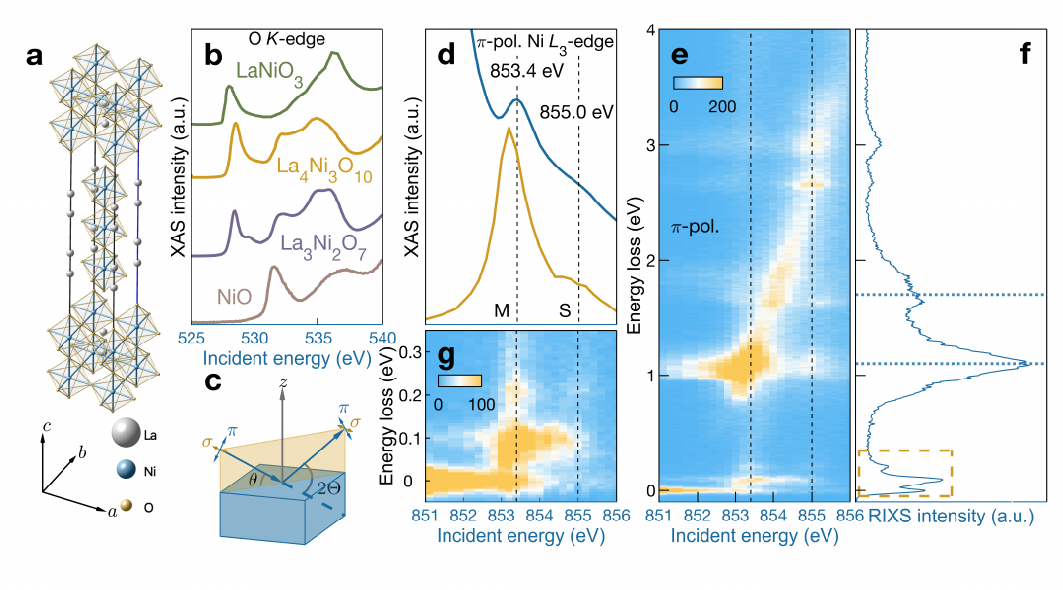}
    \caption{\textbf{XAS and incident energy dependent RIXS.} \textbf{a} Crystal structure of \LNOftt. \textbf{b} Oxygen $K$-edge XAS of NiO (pink)~\cite{chen_electronic_2024}, \LNOfts\  (purple)~\cite{chen_electronic_2024}, \LNOftt\  (yellow), and LaNiO$_3$ (green)~\cite{hepting_electronic_2020}. \textbf{c} Schematic illustration of RIXS scattering geometry. \textbf{d} Enlarged XAS spectrum (blue) measured with $\pi$-polarised X-rays at $\theta = 20^{\circ}$, overplotted with the RIXS intensity (yellow) integrated across energy loss range $\left[-0.1,0.3\right]$~eV.  \textbf{e} RIXS energy map at $\mathbf{Q} = (0.4,0)$\ measured with $\pi$-polarised X-rays. \textbf{f} Integrated RIXS spectrum across incident energy range $\left[853,854\right]$~eV of panel \textbf{e}. \textbf{g} RIXS intensity map within $\left[-0.1,0.3\right]$~eV energy loss. Vertical dashed lines mark the energies of the main and satellite resonances.}
    \label{fig:fig1}
\end{figure*}

In contrast to the well-characterized magnetic excitations in La$_3$Ni$_2$O$_{7}$, the spin dynamics in trilayer nickelates remain largely unexplored. While recent RIXS studies have reported evidence of spin excitations in trilayer Nd$_4$Ni$_3$O$_{10}$ and La$_4$Ni$_3$O$_{10}$~\cite{tenhuisen_magnetic_2025,zhang_distinct_2025}, essential characteristics including the dispersion of the excitation modes have yet to be established. As a result, important questions remain open, particularly regarding the connection between magnetic interactions and $T_c$.
This incomplete understanding of magnetism in trilayer nickelates hinders efforts to elucidate the mechanism of nickelate superconductivity.

Here, we employ high-resolution RIXS at Ni $L$-edge to study low-energy excitations in bulk La$_4$Ni$_3$O$_{10}$. 
We observe two localized spin-flip excitations at about 100 and 200 meV. In addition, dispersive magnetic excitations --- with 60~meV zone boundary energies --- are identified, which carry a comparable bandwidth but significantly reduced spectral weight compared to the collective magnetic excitations in its bilayer counterpart La$_3$Ni$_2$O$_7$.
These features are consistent with a spin-density-wave (SDW) order driven by Fermi surface nesting, suggesting a weaker electronic correlation in the trilayer system.  
Modeling the collective excitations using linear spin-wave theory with a presumed incommensurate magnetic structure yields effective intra- and inter-layer exchange parameters on similar energy scales, with the inter-layer coupling $SJ_{\perp}\approx20$~meV being the strongest, pointing to enhanced three-dimensional magnetism in La$_4$Ni$_3$O$_{10}$.
Compared to bilayer compounds, the smaller inter-layer exchange interaction is congruent with the lower $T_c$ of trilayer nickelates.
The marked differences in their magnetic excitations underscore the importance of correlation strength and dimensionality effects that should be considered when interpreting the superconducting mechanism.\\

\begin{figure*}[t]
    \centering
       \includegraphics[width=.95\textwidth]{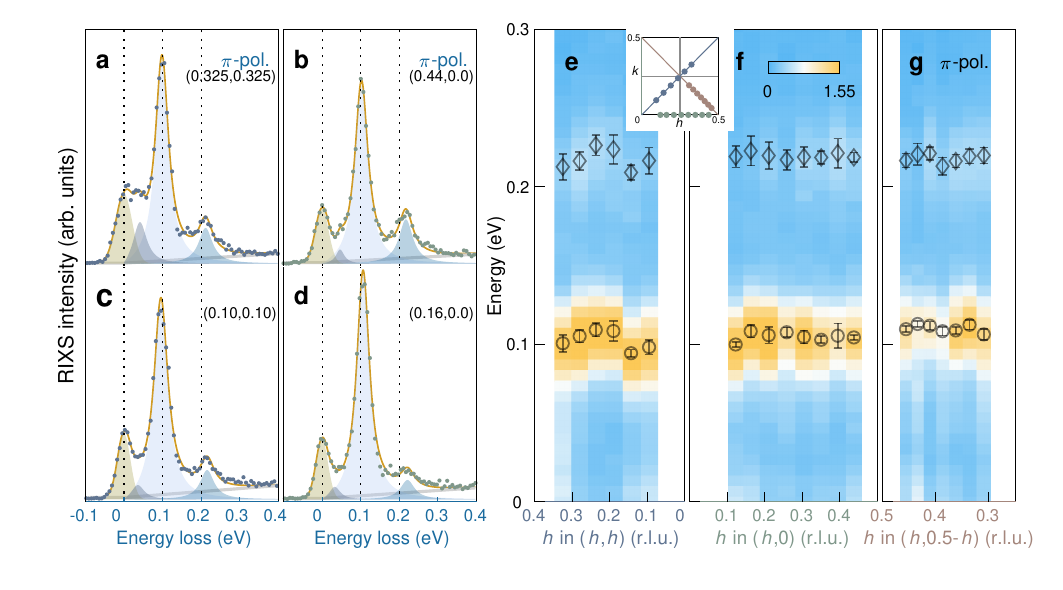}

    \caption{\textbf{Low-energy excitations and their spectral components.} \textbf{a}-\textbf{d} Representative RIXS spectra and the fits. The yellow fitting curve are composed of three components: elastic scattering (dark yellow), magnetic excitations (shades of blue), and high-energy background (gray).
    \textbf{e}-\textbf{g} RIXS intensity maps under variations of momentum and energy loss measured by $\pi$-polarized light following three paths in the reciprocal space labeled with circles in the inset. Data were taken at 23 K. Error bars represent one standard deviation.}
    \label{fig:fig2}
\end{figure*}

\begin{figure}[thb]
    \centering
    \includegraphics[width=0.5\textwidth]{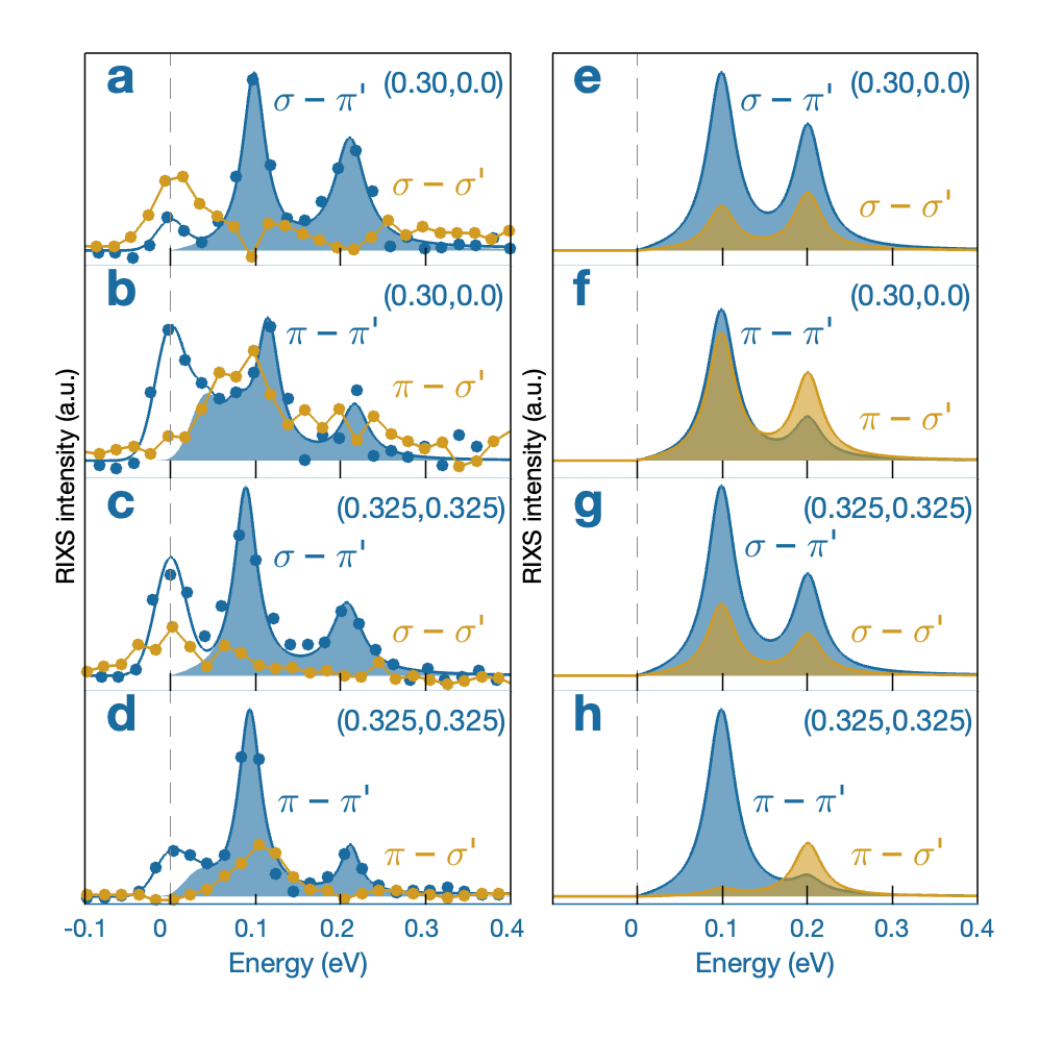}
    \caption{\textbf{Polarimetric RIXS and exact diagonalization calculations.} \textbf{a-d} Polarimetric RIXS spectra recorded with in-plane momenta as indicated with $\sigma$- and $\pi$- incident X-rays. All spectra were decomposed into components of $\sigma$-out (yellow) and $\pi$-out (blue) channels. \textbf{e-h} ED calculation results for in-plane momentum transfers and incident X-ray polarization as indicated. A life-time broadening is applied to mimic the experimental damped harmonic oscillator profile.
    }
    \label{fig:fig3}
\end{figure}

\begin{figure*}[t]
    \centering
       \includegraphics[width=.95\textwidth]{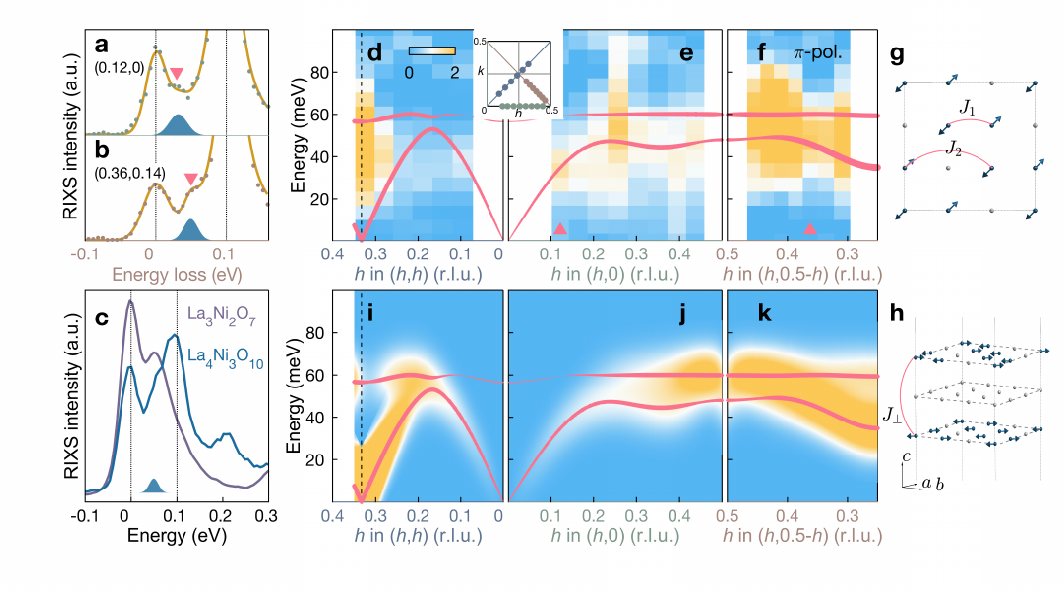}
    \caption{\textbf{Dispersion and spectral weight of the collective magnetic excitations.} \textbf{a}-\textbf{b} Selected low-energy RIXS spectra --- for momenta as indicated --- with fitted yellow curve and low-energy gaussian component filled by shades of blue. \textbf{c} Normalized integrated RIXS intensities of \LNOfts\ (purple) and \LNOftt\ (blue) with the indication of scaled gaussian component from panel \textbf{b} filled in blue~\cite{chen_electronic_2024}. \textbf{d}-\textbf{f} Low-energy RIXS intensity map with components of elastic, $0.1$~eV, and $0.2$~eV excitations subtracted. A Gaussian smoothing has been applied. The pink lines are simulated dispersion using LSWT. Downward triangles indicate the position of spectra in panels \textbf{a} and \textbf{b}. \textbf{g},\textbf{h} Magnetic structure adopted for the LSWT calculation in \textbf{i}-\textbf{k}. Blue and gray circles denote the Ni sites with and without ordered moment, respectively. The arrows indicate the spin orientations.}
    \label{fig:fig4}
\end{figure*}

\noindent\textbf{Results}\\
To investigate the electronic structure of \LNOftt, we first measure its oxygen $K$-edge X-ray absorption spectrum (XAS) and compare with those of LaNiO$_3$, \LNOfts\ and NiO~\cite{chen_electronic_2024} (Fig.~\href{fig:fig1}{1\textbf{b}}). Consistent with previous reports~\cite{zhang_distinct_2025}, a strong pre-edge peak is observed in La$_4$Ni$_3$O$_{10}$, which is also present in La$_3$Ni$_2$O$_{7}$ and LaNiO$_3$, but absent in NiO, characterizing the existence of ligand holes in RP nickelates. Fig.~\href{fig:fig1}{1\textbf{d}} displays the Ni $L_3$-edge XAS of \LNOftt, measured with $\pi$-polarization. 
Despite the partial overlap with the La $M_4$-edge (see Supplementary Fig.~\href{fig:figS1}{~1\textbf{b}}), two Ni absorption edges are identified and further confirmed by the integrated RIXS intensity in the low-energy loss range ---
shown in Fig.~\href{fig:fig1}{1\textbf{g}}.
The main resonance at 853.4~eV (labeled \textit{M}) originates from the Ni $2p\rightarrow3d^8$ excitations, while the higher-energy satellite feature at $855.0$~eV (labeled \textit{S}) corresponds to transitions to excited states with a more pronounced ligand hole character and thereby less efficient core-hole screening. This observation highlights the strong hybridization between the O $2p$ and Ni $3d$ states.

Figure \href{fig:fig1}{1\textbf{e}} displays the incident energy dependence of RIXS intensities across Ni $L_3$ resonance measured at 23~K.
Fluorescence-like signals are observed between $1$~eV and $4$~eV, originating from particle-hole excitations within the Ni-O charge continuum.
Within this energy range, $dd$ excitations are identified and selectively enhanced at the main and satellite resonances.
The integrated intensity across the main resonance energy in Fig.~\href{fig:fig1}{1\textbf{f}} reveals rich orbital excitations, with the most pounced features appearing around 1.0 and 1.7~eV, corresponding respectively to the $t_{2g}$-to-$e_g$ $dd$ excitations, and transitions involving multiple $d$ orbitals, as commonly observed in RP nickelates~\cite{chen_electronic_2024,tenhuisen_magnetic_2025}.

At lower energy, prominent excitations are observed around $\sim$$0.1$~eV,
at both the main and satellite absorption peaks (Fig.~\href{fig:fig1}{1\textbf{e}-\textbf{g}}). Their spectral weight, integrated over $[-0.1,0.3]$ eV, follows the trend of the Ni $L_3$ XAS (Fig.~\href{fig:fig1}{1\textbf{d}}), indicating a resonant electronic origin.
A closer inspection of the low-energy spectra at the main resonance (Fig.~\href{fig:fig2}{2\textbf{a}}) reveals a rich set of excitations, including a pronounced sharp mode near 100~meV, a second distinct mode at about 200~meV, and a weaker excitation around 60~meV.

Next, we investigated the momentum dependence of the low-energy excitations at the main resonance energy (853.4~eV).
The RIXS intensity as a function of energy loss and in-plane momentum transfer along high-symmetry directions reveals negligible dispersion for the excitations near 100~meV and 200~meV (Fig. \href{fig:fig2}{2\textbf{e}-\textbf{g}}). For quantitative analysis, we fitted the low-energy line shape using five components: a Gaussian profile for the elastic line, three damped harmonic oscillators (DHOs)~\cite{lamsal_extracting_2016,yan_persistent_2025} for the excitation modes, and a linear function for the background.

The absence of dispersion and resolution-limited linewidth of the $\sim$$100$- and $200$-meV modes reveal their localized characters. 
To further elucidate their origin, we performed polarimetric RIXS measurements at selective wave vectors to decompose the polarization of the outgoing X-rays. As revealed in Fig. \href{fig:fig3}{3\textbf{a}-\textbf{d}}, negligible spectral weight near $100$ and $200$~meV appears in the $\sigma-\sigma'$ channel, indicating a predominantly magnetic origin, as will be discussed later.

Although much weaker than the $\sim$$100$-meV mode, the excitation near 60 meV is clearly resolved in the raw spectra thanks to the high energy resolution, and is found to be dispersive --- see Fig. \href{fig:fig4}{4\textbf{a,b}}. To determine its dispersion, we subtract the fitted contributions of other spectral components and plot the resulting intensity as a function of energy and momentum in Fig. \href{fig:fig4}{4\textbf{d}-\textbf{f}}.
The intensity exhibits a strong modulation that increases away from the $\Gamma$ point and toward the magnetic ordering wave vector $\mathbf{Q}\mathrm{_{spin}}=(0.31,0.31)$ determined in a previous neutron diffraction study~\cite{zhang_intertwined_2020}, corroborating the propagating nature of this excitation.
This assignment is also consistent with a recent O $K$-edge RIXS and Raman spectroscopy study on La$_4$Ni$_3$O$_{10}$~\cite{zhang_distinct_2025}, which identified a bimagnon mode at an energy approximately twice that of the dispersive mode observed here.
\\

\noindent\textbf{Discussion}\\
Previous X-ray diffraction (XRD) measurements have shown that NiO$_6$ octahedra in the inner layer of trilayer nickelates possess more isotropic Ni-O bond lengths than those in the bilayer compounds~\cite{shi_absence_2025,zhu_superconductivity_2024}. In line with this structural trend, a recent transport study revealed three-dimensional superconductivity in trilayer La$_4$Ni$_3$O$_{10}$ and Pr$_4$Ni$_3$O$_{10}$, evidenced by the extremely small anisotropy of the upper critical field, which is in stark contrast with bilayer nickelates and cuprates~\cite{pei_weakly_2025}.
Considering the dominant $3 d^{8}\underline{L}$ (where $\underline{L}$ denotes a ligand hole) character of the Ni ions, we model the polarimetric RIXS results by considering a single Ni$^{2+}$ ion under a three-dimensional effective exchange field. As shown in Fig.~\href{fig:fig3}{3\textbf{e-h}}, exact diagonalization (ED) calculations assuming an effective exchange field of $J^*\approx45$~meV qualitatively reproduced both localized excitations (see Supplementary Information for details). Within this picture, the $0.1$~eV and $0.2$~eV modes originate from the local dipolar and quadrupolar spin excitations, respectively. 
This assignment accounts for the two-fold excitation energy of the $0.2$~eV mode and is consistent with the observed polarization dependence at different wave vectors.

We hypothesize that the localized excitations originate from Ni ions adopting a high-spin configuration but without ordered moment. Indeed, a previous neutron diffraction study on La$_4$Ni$_3$O$_{10}$ proposed a magnetic structure in which the inner layer is nonmagnetic~\cite{zhang_intertwined_2020}. Furthermore, the inner-layer NiO$_6$ octahedra deviate only slightly from the ideal octahedral geometry~\cite{shi_absence_2025,zhu_superconductivity_2024}, compared with those in the bilayer structure. These observations justify our interpretation of the localized spin excitations. 

To describe the dispersive low-energy magnetic excitations, we adopt a Heisenberg spin Hamiltonian motivated by earlier neutron diffraction results, which revealed an incommensurate in-plane ordering wave vector $\mathbf{Q}\mathrm{_{spin}}=(0.31,0.31)$, close to $(1/3,1/3)$, with a node at the inner layer. We therefore consider a magnetic structure consistent with this pattern, shown in Fig.~\href{fig:fig4}{4\textbf{g},\textbf{h}}, and include two in-plane exchange couplings, $J_1$ and $J_2$, and an interlayer coupling $J_\perp$.
The linear-spin-wave-theory (LSWT) calculations (Fig.~\href{fig:fig4}{4\textbf{i}-\textbf{k}}) generally capture the observed dispersion and the momentum-dependent intensity modulation, with $SJ_1=12$~meV, $SJ_2=8$~meV, and $SJ_\perp = 20$~meV providing the best agreement with the experiment.

The intensity of the dispersive magnetic excitations is nearly an order of magnitude weaker compared to that observed in La$_3$Ni$_2$O$_{7}$~\cite{chen_electronic_2024} (Fig.~\href{fig:fig4}{4\textbf{c}}). The marked suppression of the spectral weight is consistent with an itinerant SDW state. 
Recent ARPES studies on La- or Pr-based trilayer nickelates have revealed Fermi-surface nesting, band folding, and band splitting related to the SDW state~\cite{yang_electronic_2026,jiang_direct_2026}. Our RIXS results thus are consistent with such an itinerant origin of the collective magnetic excitations. Moreover, ARPES measurements also revealed significant orbital selective correlation effects in trilayer nickelates~\cite{li_orbital-selective_2026}. Under the combined influence of orbital selective electronic correlations and nesting induced instability, a complex magnetic ground state~\cite{zhang_intertwined_2020} and excitation spectrum are therefore expected, consistent with the observed coexistence of localized and collective magnetic excitations.

Compared with bilayer La$_3$Ni$_2$O$_7$, the collective magnetic excitations in La$_4$Ni$_3$O$_{10}$ exhibit a comparable bandwidth but a more pronounced three-dimensional character, as demonstrated by intra- and interlayer exchange couplings of comparable magnitude.
This is consistent with the more three-dimensional electronic structure reported for trilayer nickelates~\cite{pei_weakly_2025}.
The remarkable differences between the magnetism of the two systems reveal the importance of three-dimensionality, correlation strength, and multiorbital characters in defining the magnetic interactions and possibly the superconducting pairing mechanism in RP nickelates.
\\

\noindent\textit{Note added.} Upon completion of this work, we became aware of an independent RIXS study of magnetic and electronic excitations in La$_4$Ni$_3$O$_{10}$~\cite{chen_dissecting_2026}. \\

\noindent\textbf{Methods}\\
\noindent\textbf{Sample synthesis and characterization}\\
\noindent \LNOftt\ single crystals were grown using the high-pressure optical floating zone method, as described in ref.~\onlinecite{zhu_superconductivity_2024}. The phase purity and single crystallinity were characterized by X-ray diffraction measurements both before and after the cleave. The magnetic phase transition was characterized by transport and susceptibility measurements.
(see Supplementary Information).\\

\noindent\textbf{RIXS experiments}\\
\noindent Ni $L$-edge RIXS experiments were carried out at the ID32 beamline at the European Synchrotron Radiation Facility (ESRF)~\cite{brookes_beamline_2018}.  
Samples were aligned \textit{ex-situ} using Laue diffraction (see Supplementary Fig.~S1), and cleaved \textit{in-situ} in the load-lock chamber.
The wave vector $\textbf{Q}$ in $(q_x, q_y, q_z)$ is defined as ($h, k, l$) = $(2\pi/a,2\pi/b,2\pi/c)$ in reciprocal lattice units (r.l.u.), where $a=b=3.86$~\AA~and $c=27.97$~\AA\ are the lattice parameters of the pseudo-tetragonal unit cell.
The scattering angle is fixed to $2\Theta = 149.5^{\circ}$. The energy resolution is characterized by the full-width-at-half-maximum (FWHM) of the elastic scattering profile of silver paste or amorphous carbon which was set to $38$~meV. 
RIXS intensities were normalized to the weight of the $dd$ excitations between 0.8~eV and 2.4~eV, as in refs.~\onlinecite{wang_charge_2021,arpaia_signature_2023} (See Supplementary Note~2). Data were collected at 23 K (30 K) at ID32 (41A1).
Direct and indirect RIXS measurements were performed consecutively to resolve outgoing X-rays in $\pi'$- and $\sigma'$-polarization channels. Under the limited reflectivities of the mirror ($R_\sigma \approx 0.137$ and $R_\pi \approx 0.0703$), to attain the same photon statistics of polarimetric measurements, an increase of from six- to ten-fold exposure time is required, compared to a direct RIXS measurement.
Additional RIXS experiments were carried out at the 41A1 beamline of National Synchrotron Radiation Research Center
(NSRRC)~\cite{singh_development_2021}, yielding consistent results from those obtained at ID32 (see Supplementary Information).\\

\noindent\textbf{Single-ion and spin-wave calculations}\\
\noindent Exact diagonalization calculations based on a single-ion model were performed using the EDRIXS package~\cite{wang_edrixs_2019} to interpret the localized excitation modes. Detailed parameters for the calculation are provided in Supplementary Note 4. 
We employ an effective Heisenberg Hamiltonian to model the dispersive low-energy excitations and evaluate the magnetic interaction strengths, including two in-plane exchange couplings $J_1$ and $J_2$ and one interlayer coupling $J_\perp$:
\begin{equation}
H = J_1 \sum_{\langle i, j \rangle} \mathbf{S}_i \cdot \mathbf{S}_j
  + J_2 \sum_{\langle i, i' \rangle} \mathbf{S}_i \cdot \mathbf{S}_{i'}
  + J_\perp \sum_{\langle i, j' \rangle} \mathbf{S}_i \cdot \mathbf{S}_{j'}
\end{equation}
where $\langle i, j \rangle$, $\langle i, i' \rangle$, and $\langle i, j' \rangle$ denote pairs of coupled spin sites as indicated in Fig.~\href{fig:fig4}{4\textbf{g},\textbf{h}}, and $\mathbf{S}_i$ denotes the spin operator at the lattice site $i$. The exchange couplings were determined by comparing the momentum dependent RIXS intensity with the calculated dynamic spin structure factor, with its modulation due to the varying $l$ component in the RIXS geometry accounted. The SpinW package was used for the spin-wave calculations~\cite{toth_linear_2015}.
\\

\noindent\textbf{Acknowledgments}\\
We thank Yi Lu and Kun Jiang for insightful discussions. The work at CUHK is supported by the Research Grants Council of Hong Kong (ECS No. 24306223), and the Guangdong Provincial Quantum Science Strategic Initiative (GDZX2401012), and the 1+1+1 CUHK-CUHK(SZ)-GDSTC Joint Collaboration Fund (Project Code: 2025A0505000079).
Y.Z., E.Z., L.C. and J.Z. acknowledge support from The Key Program of the National Natural Science Foundation of China (Grant No. 12234006), the National Key R\&D Program of China (Grant No. 2022YFA1403202), the Quantum Science and Technology-National Science and Technology Major Project
(Grant No. 2024ZD0300103), the Shanghai Municipal Science and Technology Project (Grants No. 2019SHZDZX01 and No. 25DZ3008100).
Y.S. acknowledges support from the National Key R\&D Program of China (Grant No.~2024YFA1408301) and the National Natural Science Foundation of China (Grant No.~12574139).
X.H. and J.C. thank the Swiss National Science Foundation under Projects No. 200021\_188564.
We acknowledge the ID32 and 41A1 Beamlines for providing beamtime under Proposals HC-5878 and 2024-1-098-1.
\\

\noindent\textbf{Author contributions}\\
Q.W. conceived the project. Y.Z, E.Z. L.C, and J.Z grew the La$_4$Ni$_{3}$O$_{10}$ single crystals, and characterized the samples with Y.C. and T.W. 
Y.C., Y.Y., X.H., M.R.C. and N.B.B. carried out the RIXS experiments at ID32. Y.C., Y.L., J.O., H.Y.H. and D.J.H carried out the RIXS experiments at 41A1. Y.C., Y. L., Y.S. and Q.W. analyzed the RIXS data. Y.C., J.C. and Q.W. wrote the manuscript with input from other authors.\\
	
\noindent\textbf{Data availability} \\
Data supporting the findings of this study are available from the corresponding authors upon request.\\
	
\noindent\textbf{Competing interests} \\
The authors declare no competing interests.\\


\begin{thebibliography}{37}%
\makeatletter
\providecommand \@ifxundefined [1]{%
 \@ifx{#1\undefined}
}%
\providecommand \@ifnum [1]{%
 \ifnum #1\expandafter \@firstoftwo
 \else \expandafter \@secondoftwo
 \fi
}%
\providecommand \@ifx [1]{%
 \ifx #1\expandafter \@firstoftwo
 \else \expandafter \@secondoftwo
 \fi
}%
\providecommand \natexlab [1]{#1}%
\providecommand \enquote  [1]{``#1''}%
\providecommand \bibnamefont  [1]{#1}%
\providecommand \bibfnamefont [1]{#1}%
\providecommand \citenamefont [1]{#1}%
\providecommand \href@noop [0]{\@secondoftwo}%
\providecommand \href [0]{\begingroup \@sanitize@url \@href}%
\providecommand \@href[1]{\@@startlink{#1}\@@href}%
\providecommand \@@href[1]{\endgroup#1\@@endlink}%
\providecommand \@sanitize@url [0]{\catcode `\\12\catcode `\$12\catcode `\&12\catcode `\#12\catcode `\^12\catcode `\_12\catcode `\%12\relax}%
\providecommand \@@startlink[1]{}%
\providecommand \@@endlink[0]{}%
\providecommand \url  [0]{\begingroup\@sanitize@url \@url }%
\providecommand \@url [1]{\endgroup\@href {#1}{\urlprefix }}%
\providecommand \urlprefix  [0]{URL }%
\providecommand \Eprint [0]{\href }%
\providecommand \doibase [0]{http://dx.doi.org/}%
\providecommand \selectlanguage [0]{\@gobble}%
\providecommand \bibinfo  [0]{\@secondoftwo}%
\providecommand \bibfield  [0]{\@secondoftwo}%
\providecommand \translation [1]{[#1]}%
\providecommand \BibitemOpen [0]{}%
\providecommand \bibitemStop [0]{}%
\providecommand \bibitemNoStop [0]{.\EOS\space}%
\providecommand \EOS [0]{\spacefactor3000\relax}%
\providecommand \BibitemShut  [1]{\csname bibitem#1\endcsname}%
\let\auto@bib@innerbib\@empty
\bibitem [{\citenamefont {Lee}\ \emph {et~al.}(2006)\citenamefont {Lee}, \citenamefont {Nagaosa},\ and\ \citenamefont {Wen}}]{lee_doping_2006}%
  \BibitemOpen
  \bibfield  {author} {\bibinfo {author} {\bibfnamefont {P.~A.}\ \bibnamefont {Lee}}, \bibinfo {author} {\bibfnamefont {N.}~\bibnamefont {Nagaosa}}, \ and\ \bibinfo {author} {\bibfnamefont {X.-G.}\ \bibnamefont {Wen}},\ }\href {\doibase 10.1103/revmodphys.78.17} {\bibfield  {journal} {\bibinfo  {journal} {Rev. Mod. Phys.}\ }\textbf {\bibinfo {volume} {78}},\ \bibinfo {pages} {17} (\bibinfo {year} {2006})}\BibitemShut {NoStop}%
\bibitem [{\citenamefont {Keimer}\ \emph {et~al.}(2015)\citenamefont {Keimer}, \citenamefont {Kivelson}, \citenamefont {Norman}, \citenamefont {Uchida},\ and\ \citenamefont {Zaanen}}]{keimer_from_2015}%
  \BibitemOpen
  \bibfield  {author} {\bibinfo {author} {\bibfnamefont {B.}~\bibnamefont {Keimer}}, \bibinfo {author} {\bibfnamefont {S.~A.}\ \bibnamefont {Kivelson}}, \bibinfo {author} {\bibfnamefont {M.~R.}\ \bibnamefont {Norman}}, \bibinfo {author} {\bibfnamefont {S.}~\bibnamefont {Uchida}}, \ and\ \bibinfo {author} {\bibfnamefont {J.}~\bibnamefont {Zaanen}},\ }\href {\doibase 10.1038/nature14165} {\bibfield  {journal} {\bibinfo  {journal} {Nature}\ }\textbf {\bibinfo {volume} {518}},\ \bibinfo {pages} {179} (\bibinfo {year} {2015})}\BibitemShut {NoStop}%
\bibitem [{\citenamefont {Sun}\ \emph {et~al.}(2023)\citenamefont {Sun}, \citenamefont {Huo}, \citenamefont {Hu}, \citenamefont {Li}, \citenamefont {Liu}, \citenamefont {Han}, \citenamefont {Tang}, \citenamefont {Mao}, \citenamefont {Yang}, \citenamefont {Wang}, \citenamefont {Cheng}, \citenamefont {Yao}, \citenamefont {Zhang},\ and\ \citenamefont {Wang}}]{sun_signatures_2023}%
  \BibitemOpen
  \bibfield  {author} {\bibinfo {author} {\bibfnamefont {H.}~\bibnamefont {Sun}}, \bibinfo {author} {\bibfnamefont {M.}~\bibnamefont {Huo}}, \bibinfo {author} {\bibfnamefont {X.}~\bibnamefont {Hu}}, \bibinfo {author} {\bibfnamefont {J.}~\bibnamefont {Li}}, \bibinfo {author} {\bibfnamefont {Z.}~\bibnamefont {Liu}}, \bibinfo {author} {\bibfnamefont {Y.}~\bibnamefont {Han}}, \bibinfo {author} {\bibfnamefont {L.}~\bibnamefont {Tang}}, \bibinfo {author} {\bibfnamefont {Z.}~\bibnamefont {Mao}}, \bibinfo {author} {\bibfnamefont {P.}~\bibnamefont {Yang}}, \bibinfo {author} {\bibfnamefont {B.}~\bibnamefont {Wang}}, \bibinfo {author} {\bibfnamefont {J.}~\bibnamefont {Cheng}}, \bibinfo {author} {\bibfnamefont {D.-X.}\ \bibnamefont {Yao}}, \bibinfo {author} {\bibfnamefont {G.-M.}\ \bibnamefont {Zhang}}, \ and\ \bibinfo {author} {\bibfnamefont {M.}~\bibnamefont {Wang}},\ }\href {\doibase 10.1038/s41586-023-06408-7} {\bibfield  {journal} {\bibinfo  {journal} {Nature}\ }\textbf {\bibinfo {volume} {621}},\ \bibinfo {pages}
  {493} (\bibinfo {year} {2023})}\BibitemShut {NoStop}%
\bibitem [{\citenamefont {Zhu}\ \emph {et~al.}(2024)\citenamefont {Zhu}, \citenamefont {Peng}, \citenamefont {Zhang}, \citenamefont {Pan}, \citenamefont {Chen}, \citenamefont {Chen}, \citenamefont {Ren}, \citenamefont {Liu}, \citenamefont {Hao}, \citenamefont {Li}, \citenamefont {Xing}, \citenamefont {Lan}, \citenamefont {Han}, \citenamefont {Wang}, \citenamefont {Jia}, \citenamefont {Wo}, \citenamefont {Gu}, \citenamefont {Gu}, \citenamefont {Ji}, \citenamefont {Wang}, \citenamefont {Gou}, \citenamefont {Shen}, \citenamefont {Ying}, \citenamefont {Chen}, \citenamefont {Yang}, \citenamefont {Cao}, \citenamefont {Zheng}, \citenamefont {Zeng}, \citenamefont {Guo},\ and\ \citenamefont {Zhao}}]{zhu_superconductivity_2024}%
  \BibitemOpen
  \bibfield  {author} {\bibinfo {author} {\bibfnamefont {Y.}~\bibnamefont {Zhu}}, \bibinfo {author} {\bibfnamefont {D.}~\bibnamefont {Peng}}, \bibinfo {author} {\bibfnamefont {E.}~\bibnamefont {Zhang}}, \bibinfo {author} {\bibfnamefont {B.}~\bibnamefont {Pan}}, \bibinfo {author} {\bibfnamefont {X.}~\bibnamefont {Chen}}, \bibinfo {author} {\bibfnamefont {L.}~\bibnamefont {Chen}}, \bibinfo {author} {\bibfnamefont {H.}~\bibnamefont {Ren}}, \bibinfo {author} {\bibfnamefont {F.}~\bibnamefont {Liu}}, \bibinfo {author} {\bibfnamefont {Y.}~\bibnamefont {Hao}}, \bibinfo {author} {\bibfnamefont {N.}~\bibnamefont {Li}}, \bibinfo {author} {\bibfnamefont {Z.}~\bibnamefont {Xing}}, \bibinfo {author} {\bibfnamefont {F.}~\bibnamefont {Lan}}, \bibinfo {author} {\bibfnamefont {J.}~\bibnamefont {Han}}, \bibinfo {author} {\bibfnamefont {J.}~\bibnamefont {Wang}}, \bibinfo {author} {\bibfnamefont {D.}~\bibnamefont {Jia}}, \bibinfo {author} {\bibfnamefont {H.}~\bibnamefont {Wo}}, \bibinfo {author} {\bibfnamefont {Y.}~\bibnamefont
  {Gu}}, \bibinfo {author} {\bibfnamefont {Y.}~\bibnamefont {Gu}}, \bibinfo {author} {\bibfnamefont {L.}~\bibnamefont {Ji}}, \bibinfo {author} {\bibfnamefont {W.}~\bibnamefont {Wang}}, \bibinfo {author} {\bibfnamefont {H.}~\bibnamefont {Gou}}, \bibinfo {author} {\bibfnamefont {Y.}~\bibnamefont {Shen}}, \bibinfo {author} {\bibfnamefont {T.}~\bibnamefont {Ying}}, \bibinfo {author} {\bibfnamefont {X.}~\bibnamefont {Chen}}, \bibinfo {author} {\bibfnamefont {W.}~\bibnamefont {Yang}}, \bibinfo {author} {\bibfnamefont {H.}~\bibnamefont {Cao}}, \bibinfo {author} {\bibfnamefont {C.}~\bibnamefont {Zheng}}, \bibinfo {author} {\bibfnamefont {Q.}~\bibnamefont {Zeng}}, \bibinfo {author} {\bibfnamefont {J.-g.}\ \bibnamefont {Guo}}, \ and\ \bibinfo {author} {\bibfnamefont {J.}~\bibnamefont {Zhao}},\ }\href {\doibase 10.1038/s41586-024-07553-3} {\bibfield  {journal} {\bibinfo  {journal} {Nature}\ }\textbf {\bibinfo {volume} {631}},\ \bibinfo {pages} {531} (\bibinfo {year} {2024})}\BibitemShut {NoStop}%
\bibitem [{\citenamefont {Li}\ \emph {et~al.}(2017)\citenamefont {Li}, \citenamefont {Zhou}, \citenamefont {Nummy}, \citenamefont {Zhang}, \citenamefont {Pardo}, \citenamefont {Pickett}, \citenamefont {Mitchell},\ and\ \citenamefont {Dessau}}]{li_fermiology_2017}%
  \BibitemOpen
  \bibfield  {author} {\bibinfo {author} {\bibfnamefont {H.}~\bibnamefont {Li}}, \bibinfo {author} {\bibfnamefont {X.}~\bibnamefont {Zhou}}, \bibinfo {author} {\bibfnamefont {T.}~\bibnamefont {Nummy}}, \bibinfo {author} {\bibfnamefont {J.}~\bibnamefont {Zhang}}, \bibinfo {author} {\bibfnamefont {V.}~\bibnamefont {Pardo}}, \bibinfo {author} {\bibfnamefont {W.~E.}\ \bibnamefont {Pickett}}, \bibinfo {author} {\bibfnamefont {J.~F.}\ \bibnamefont {Mitchell}}, \ and\ \bibinfo {author} {\bibfnamefont {D.~S.}\ \bibnamefont {Dessau}},\ }\href {\doibase 10.1038/s41467-017-00777-0} {\bibfield  {journal} {\bibinfo  {journal} {Nat. Commun.}\ }\textbf {\bibinfo {volume} {8}},\ \bibinfo {pages} {704} (\bibinfo {year} {2017})}\BibitemShut {NoStop}%
\bibitem [{\citenamefont {Yang}\ \emph {et~al.}(2024)\citenamefont {Yang}, \citenamefont {Sun}, \citenamefont {Hu}, \citenamefont {Xie}, \citenamefont {Miao}, \citenamefont {Luo}, \citenamefont {Chen}, \citenamefont {Liang}, \citenamefont {Zhu}, \citenamefont {Qu}, \citenamefont {Chen}, \citenamefont {Huo}, \citenamefont {Huang}, \citenamefont {Zhang}, \citenamefont {Zhang}, \citenamefont {Yang}, \citenamefont {Wang}, \citenamefont {Peng}, \citenamefont {Mao}, \citenamefont {Liu}, \citenamefont {Xu}, \citenamefont {Qian}, \citenamefont {Yao}, \citenamefont {Wang}, \citenamefont {Zhao},\ and\ \citenamefont {Zhou}}]{yang_orbital_2024}%
  \BibitemOpen
  \bibfield  {author} {\bibinfo {author} {\bibfnamefont {J.}~\bibnamefont {Yang}}, \bibinfo {author} {\bibfnamefont {H.}~\bibnamefont {Sun}}, \bibinfo {author} {\bibfnamefont {X.}~\bibnamefont {Hu}}, \bibinfo {author} {\bibfnamefont {Y.}~\bibnamefont {Xie}}, \bibinfo {author} {\bibfnamefont {T.}~\bibnamefont {Miao}}, \bibinfo {author} {\bibfnamefont {H.}~\bibnamefont {Luo}}, \bibinfo {author} {\bibfnamefont {H.}~\bibnamefont {Chen}}, \bibinfo {author} {\bibfnamefont {B.}~\bibnamefont {Liang}}, \bibinfo {author} {\bibfnamefont {W.}~\bibnamefont {Zhu}}, \bibinfo {author} {\bibfnamefont {G.}~\bibnamefont {Qu}}, \bibinfo {author} {\bibfnamefont {C.-Q.}\ \bibnamefont {Chen}}, \bibinfo {author} {\bibfnamefont {M.}~\bibnamefont {Huo}}, \bibinfo {author} {\bibfnamefont {Y.}~\bibnamefont {Huang}}, \bibinfo {author} {\bibfnamefont {S.}~\bibnamefont {Zhang}}, \bibinfo {author} {\bibfnamefont {F.}~\bibnamefont {Zhang}}, \bibinfo {author} {\bibfnamefont {F.}~\bibnamefont {Yang}}, \bibinfo {author} {\bibfnamefont
  {Z.}~\bibnamefont {Wang}}, \bibinfo {author} {\bibfnamefont {Q.}~\bibnamefont {Peng}}, \bibinfo {author} {\bibfnamefont {H.}~\bibnamefont {Mao}}, \bibinfo {author} {\bibfnamefont {G.}~\bibnamefont {Liu}}, \bibinfo {author} {\bibfnamefont {Z.}~\bibnamefont {Xu}}, \bibinfo {author} {\bibfnamefont {T.}~\bibnamefont {Qian}}, \bibinfo {author} {\bibfnamefont {D.-X.}\ \bibnamefont {Yao}}, \bibinfo {author} {\bibfnamefont {M.}~\bibnamefont {Wang}}, \bibinfo {author} {\bibfnamefont {L.}~\bibnamefont {Zhao}}, \ and\ \bibinfo {author} {\bibfnamefont {X.~J.}\ \bibnamefont {Zhou}},\ }\href {\doibase 10.1038/s41467-024-48701-7} {\bibfield  {journal} {\bibinfo  {journal} {Nat. Commun.}\ }\textbf {\bibinfo {volume} {15}},\ \bibinfo {pages} {4373} (\bibinfo {year} {2024})}\BibitemShut {NoStop}%
\bibitem [{\citenamefont {Au-Yeung}\ \emph {et~al.}(2025)\citenamefont {Au-Yeung}, \citenamefont {Chen}, \citenamefont {Smit}, \citenamefont {Bluschke}, \citenamefont {Zimmermann}, \citenamefont {Michiardi}, \citenamefont {Moen}, \citenamefont {Kraan}, \citenamefont {Pang}, \citenamefont {Suen}, \citenamefont {Zhdanovich}, \citenamefont {Zonno}, \citenamefont {Gorovikov}, \citenamefont {Liu}, \citenamefont {Levy}, \citenamefont {Elfimov}, \citenamefont {Berciu}, \citenamefont {Sawatzky}, \citenamefont {Mitchell},\ and\ \citenamefont {Damascelli}}]{au-yeung_universal_2025}%
  \BibitemOpen
  \bibfield  {author} {\bibinfo {author} {\bibfnamefont {C.~C.}\ \bibnamefont {Au-Yeung}}, \bibinfo {author} {\bibfnamefont {X.}~\bibnamefont {Chen}}, \bibinfo {author} {\bibfnamefont {S.}~\bibnamefont {Smit}}, \bibinfo {author} {\bibfnamefont {M.}~\bibnamefont {Bluschke}}, \bibinfo {author} {\bibfnamefont {V.}~\bibnamefont {Zimmermann}}, \bibinfo {author} {\bibfnamefont {M.}~\bibnamefont {Michiardi}}, \bibinfo {author} {\bibfnamefont {P.~C.}\ \bibnamefont {Moen}}, \bibinfo {author} {\bibfnamefont {J.}~\bibnamefont {Kraan}}, \bibinfo {author} {\bibfnamefont {C.~S.~B.}\ \bibnamefont {Pang}}, \bibinfo {author} {\bibfnamefont {C.~T.}\ \bibnamefont {Suen}}, \bibinfo {author} {\bibfnamefont {S.}~\bibnamefont {Zhdanovich}}, \bibinfo {author} {\bibfnamefont {M.}~\bibnamefont {Zonno}}, \bibinfo {author} {\bibfnamefont {S.}~\bibnamefont {Gorovikov}}, \bibinfo {author} {\bibfnamefont {Y.}~\bibnamefont {Liu}}, \bibinfo {author} {\bibfnamefont {G.}~\bibnamefont {Levy}}, \bibinfo {author} {\bibfnamefont {I.~S.}\
  \bibnamefont {Elfimov}}, \bibinfo {author} {\bibfnamefont {M.}~\bibnamefont {Berciu}}, \bibinfo {author} {\bibfnamefont {G.~A.}\ \bibnamefont {Sawatzky}}, \bibinfo {author} {\bibfnamefont {J.~F.}\ \bibnamefont {Mitchell}}, \ and\ \bibinfo {author} {\bibfnamefont {A.}~\bibnamefont {Damascelli}},\ }\href {http://arxiv.org/abs/2502.20450} {\bibfield  {journal} {\bibinfo  {journal} {arXiv:2502.20450}\ } (\bibinfo {year} {2025})}\BibitemShut {NoStop}%
\bibitem [{\citenamefont {Li}\ \emph {et~al.}(2025)\citenamefont {Li}, \citenamefont {Zhou}, \citenamefont {Lv}, \citenamefont {Li}, \citenamefont {Yue}, \citenamefont {Huang}, \citenamefont {Xu}, \citenamefont {Shen}, \citenamefont {Miao}, \citenamefont {Song}, \citenamefont {Nie}, \citenamefont {Chen}, \citenamefont {Wang}, \citenamefont {Chen}, \citenamefont {Huang}, \citenamefont {Chen}, \citenamefont {Qian}, \citenamefont {Lin}, \citenamefont {He}, \citenamefont {Sun}, \citenamefont {Chen},\ and\ \citenamefont {Xue}}]{li_angle-resolved_2025}%
  \BibitemOpen
  \bibfield  {author} {\bibinfo {author} {\bibfnamefont {P.}~\bibnamefont {Li}}, \bibinfo {author} {\bibfnamefont {G.}~\bibnamefont {Zhou}}, \bibinfo {author} {\bibfnamefont {W.}~\bibnamefont {Lv}}, \bibinfo {author} {\bibfnamefont {Y.}~\bibnamefont {Li}}, \bibinfo {author} {\bibfnamefont {C.}~\bibnamefont {Yue}}, \bibinfo {author} {\bibfnamefont {H.}~\bibnamefont {Huang}}, \bibinfo {author} {\bibfnamefont {L.}~\bibnamefont {Xu}}, \bibinfo {author} {\bibfnamefont {J.}~\bibnamefont {Shen}}, \bibinfo {author} {\bibfnamefont {Y.}~\bibnamefont {Miao}}, \bibinfo {author} {\bibfnamefont {W.}~\bibnamefont {Song}}, \bibinfo {author} {\bibfnamefont {Z.}~\bibnamefont {Nie}}, \bibinfo {author} {\bibfnamefont {Y.}~\bibnamefont {Chen}}, \bibinfo {author} {\bibfnamefont {H.}~\bibnamefont {Wang}}, \bibinfo {author} {\bibfnamefont {W.}~\bibnamefont {Chen}}, \bibinfo {author} {\bibfnamefont {Y.}~\bibnamefont {Huang}}, \bibinfo {author} {\bibfnamefont {Z.-H.}\ \bibnamefont {Chen}}, \bibinfo {author} {\bibfnamefont
  {T.}~\bibnamefont {Qian}}, \bibinfo {author} {\bibfnamefont {J.}~\bibnamefont {Lin}}, \bibinfo {author} {\bibfnamefont {J.}~\bibnamefont {He}}, \bibinfo {author} {\bibfnamefont {Y.-J.}\ \bibnamefont {Sun}}, \bibinfo {author} {\bibfnamefont {Z.}~\bibnamefont {Chen}}, \ and\ \bibinfo {author} {\bibfnamefont {Q.-K.}\ \bibnamefont {Xue}},\ }\href {\doibase 10.1093/nsr/nwaf205} {\bibfield  {journal} {\bibinfo  {journal} {Natl. Sci. Rev.}\ }\textbf {\bibinfo {volume} {12}},\ \bibinfo {pages} {nwaf205} (\bibinfo {year} {2025})}\BibitemShut {NoStop}%
\bibitem [{\citenamefont {Chen}\ \emph {et~al.}(2024)\citenamefont {Chen}, \citenamefont {Choi}, \citenamefont {Jiang}, \citenamefont {Mei}, \citenamefont {Jiang}, \citenamefont {Li}, \citenamefont {Agrestini}, \citenamefont {Garcia-Fernandez}, \citenamefont {Sun}, \citenamefont {Huang}, \citenamefont {Shen}, \citenamefont {Wang}, \citenamefont {Hu}, \citenamefont {Lu}, \citenamefont {Zhou},\ and\ \citenamefont {Feng}}]{chen_electronic_2024}%
  \BibitemOpen
  \bibfield  {author} {\bibinfo {author} {\bibfnamefont {X.}~\bibnamefont {Chen}}, \bibinfo {author} {\bibfnamefont {J.}~\bibnamefont {Choi}}, \bibinfo {author} {\bibfnamefont {Z.}~\bibnamefont {Jiang}}, \bibinfo {author} {\bibfnamefont {J.}~\bibnamefont {Mei}}, \bibinfo {author} {\bibfnamefont {K.}~\bibnamefont {Jiang}}, \bibinfo {author} {\bibfnamefont {J.}~\bibnamefont {Li}}, \bibinfo {author} {\bibfnamefont {S.}~\bibnamefont {Agrestini}}, \bibinfo {author} {\bibfnamefont {M.}~\bibnamefont {Garcia-Fernandez}}, \bibinfo {author} {\bibfnamefont {H.}~\bibnamefont {Sun}}, \bibinfo {author} {\bibfnamefont {X.}~\bibnamefont {Huang}}, \bibinfo {author} {\bibfnamefont {D.}~\bibnamefont {Shen}}, \bibinfo {author} {\bibfnamefont {M.}~\bibnamefont {Wang}}, \bibinfo {author} {\bibfnamefont {J.}~\bibnamefont {Hu}}, \bibinfo {author} {\bibfnamefont {Y.}~\bibnamefont {Lu}}, \bibinfo {author} {\bibfnamefont {K.-J.}\ \bibnamefont {Zhou}}, \ and\ \bibinfo {author} {\bibfnamefont {D.}~\bibnamefont {Feng}},\ }\href {\doibase
  10.1038/s41467-024-53863-5} {\bibfield  {journal} {\bibinfo  {journal} {Nat. Commun.}\ }\textbf {\bibinfo {volume} {15}},\ \bibinfo {pages} {9597} (\bibinfo {year} {2024})}\BibitemShut {NoStop}%
\bibitem [{\citenamefont {Zhong}\ \emph {et~al.}(2025)\citenamefont {Zhong}, \citenamefont {Hao}, \citenamefont {Zhang}, \citenamefont {Chen}, \citenamefont {Wei}, \citenamefont {Liu}, \citenamefont {Huang}, \citenamefont {Li}, \citenamefont {Zhang}, \citenamefont {Liu}, \citenamefont {Ni}, \citenamefont {dos Reis~Cantarino}, \citenamefont {Kummer}, \citenamefont {Brookes}, \citenamefont {Cao}, \citenamefont {Nie}, \citenamefont {Schmitt},\ and\ \citenamefont {Lu}}]{zhong_spin_2025}%
  \BibitemOpen
  \bibfield  {author} {\bibinfo {author} {\bibfnamefont {H.}~\bibnamefont {Zhong}}, \bibinfo {author} {\bibfnamefont {B.}~\bibnamefont {Hao}}, \bibinfo {author} {\bibfnamefont {Z.}~\bibnamefont {Zhang}}, \bibinfo {author} {\bibfnamefont {A.}~\bibnamefont {Chen}}, \bibinfo {author} {\bibfnamefont {Y.}~\bibnamefont {Wei}}, \bibinfo {author} {\bibfnamefont {R.}~\bibnamefont {Liu}}, \bibinfo {author} {\bibfnamefont {X.}~\bibnamefont {Huang}}, \bibinfo {author} {\bibfnamefont {C.}~\bibnamefont {Li}}, \bibinfo {author} {\bibfnamefont {W.}~\bibnamefont {Zhang}}, \bibinfo {author} {\bibfnamefont {C.}~\bibnamefont {Liu}}, \bibinfo {author} {\bibfnamefont {X.-S.}\ \bibnamefont {Ni}}, \bibinfo {author} {\bibfnamefont {M.}~\bibnamefont {dos Reis~Cantarino}}, \bibinfo {author} {\bibfnamefont {K.}~\bibnamefont {Kummer}}, \bibinfo {author} {\bibfnamefont {N.}~\bibnamefont {Brookes}}, \bibinfo {author} {\bibfnamefont {K.}~\bibnamefont {Cao}}, \bibinfo {author} {\bibfnamefont {Y.}~\bibnamefont {Nie}}, \bibinfo {author}
  {\bibfnamefont {T.}~\bibnamefont {Schmitt}}, \ and\ \bibinfo {author} {\bibfnamefont {X.}~\bibnamefont {Lu}},\ }\href {https://arxiv.org/abs/2502.03178} {\bibfield  {journal} {\bibinfo  {journal} {arXiv:2502.03178}\ } (\bibinfo {year} {2025})}\BibitemShut {NoStop}%
\bibitem [{\citenamefont {Xie}\ \emph {et~al.}(2024)\citenamefont {Xie}, \citenamefont {Huo}, \citenamefont {Ni}, \citenamefont {Shen}, \citenamefont {Huang}, \citenamefont {Sun}, \citenamefont {Walker}, \citenamefont {Adroja}, \citenamefont {Yu}, \citenamefont {Shen}, \citenamefont {He}, \citenamefont {Cao},\ and\ \citenamefont {Wang}}]{xie_strong_2024}%
  \BibitemOpen
  \bibfield  {author} {\bibinfo {author} {\bibfnamefont {T.}~\bibnamefont {Xie}}, \bibinfo {author} {\bibfnamefont {M.}~\bibnamefont {Huo}}, \bibinfo {author} {\bibfnamefont {X.}~\bibnamefont {Ni}}, \bibinfo {author} {\bibfnamefont {F.}~\bibnamefont {Shen}}, \bibinfo {author} {\bibfnamefont {X.}~\bibnamefont {Huang}}, \bibinfo {author} {\bibfnamefont {H.}~\bibnamefont {Sun}}, \bibinfo {author} {\bibfnamefont {H.~C.}\ \bibnamefont {Walker}}, \bibinfo {author} {\bibfnamefont {D.}~\bibnamefont {Adroja}}, \bibinfo {author} {\bibfnamefont {D.}~\bibnamefont {Yu}}, \bibinfo {author} {\bibfnamefont {B.}~\bibnamefont {Shen}}, \bibinfo {author} {\bibfnamefont {L.}~\bibnamefont {He}}, \bibinfo {author} {\bibfnamefont {K.}~\bibnamefont {Cao}}, \ and\ \bibinfo {author} {\bibfnamefont {M.}~\bibnamefont {Wang}},\ }\href {\doibase 10.1016/j.scib.2024.07.030} {\bibfield  {journal} {\bibinfo  {journal} {Sci. Bull.}\ }\textbf {\bibinfo {volume} {69}},\ \bibinfo {pages} {3221} (\bibinfo {year} {2024})}\BibitemShut {NoStop}%
\bibitem [{\citenamefont {Liu}\ \emph {et~al.}(2023)\citenamefont {Liu}, \citenamefont {Mei}, \citenamefont {Ye}, \citenamefont {Chen},\ and\ \citenamefont {Yang}}]{liu_swave_2023}%
  \BibitemOpen
  \bibfield  {author} {\bibinfo {author} {\bibfnamefont {Y.-B.}\ \bibnamefont {Liu}}, \bibinfo {author} {\bibfnamefont {J.-W.}\ \bibnamefont {Mei}}, \bibinfo {author} {\bibfnamefont {F.}~\bibnamefont {Ye}}, \bibinfo {author} {\bibfnamefont {W.-Q.}\ \bibnamefont {Chen}}, \ and\ \bibinfo {author} {\bibfnamefont {F.}~\bibnamefont {Yang}},\ }\href {\doibase 10.1103/PhysRevLett.131.236002} {\bibfield  {journal} {\bibinfo  {journal} {Phys. Rev. Lett.}\ }\textbf {\bibinfo {volume} {131}},\ \bibinfo {pages} {236002} (\bibinfo {year} {2023})}\BibitemShut {NoStop}%
\bibitem [{\citenamefont {Yang}\ \emph {et~al.}(2023)\citenamefont {Yang}, \citenamefont {Wang},\ and\ \citenamefont {Wang}}]{yang_possible_2023}%
  \BibitemOpen
  \bibfield  {author} {\bibinfo {author} {\bibfnamefont {Q.-G.}\ \bibnamefont {Yang}}, \bibinfo {author} {\bibfnamefont {D.}~\bibnamefont {Wang}}, \ and\ \bibinfo {author} {\bibfnamefont {Q.-H.}\ \bibnamefont {Wang}},\ }\href {\doibase 10.1103/PhysRevB.108.L140505} {\bibfield  {journal} {\bibinfo  {journal} {Phys. Rev. B}\ }\textbf {\bibinfo {volume} {108}},\ \bibinfo {pages} {L140505} (\bibinfo {year} {2023})}\BibitemShut {NoStop}%
\bibitem [{\citenamefont {Zhang}\ \emph {et~al.}(2023)\citenamefont {Zhang}, \citenamefont {Lin}, \citenamefont {Moreo}, \citenamefont {Maier},\ and\ \citenamefont {Dagotto}}]{zhang_trends_2023}%
  \BibitemOpen
  \bibfield  {author} {\bibinfo {author} {\bibfnamefont {Y.}~\bibnamefont {Zhang}}, \bibinfo {author} {\bibfnamefont {L.-F.}\ \bibnamefont {Lin}}, \bibinfo {author} {\bibfnamefont {A.}~\bibnamefont {Moreo}}, \bibinfo {author} {\bibfnamefont {T.~A.}\ \bibnamefont {Maier}}, \ and\ \bibinfo {author} {\bibfnamefont {E.}~\bibnamefont {Dagotto}},\ }\href {\doibase 10.1103/PhysRevB.108.165141} {\bibfield  {journal} {\bibinfo  {journal} {Phys. Rev. B}\ }\textbf {\bibinfo {volume} {108}},\ \bibinfo {pages} {165141} (\bibinfo {year} {2023})}\BibitemShut {NoStop}%
\bibitem [{\citenamefont {Lechermann}\ \emph {et~al.}(2023)\citenamefont {Lechermann}, \citenamefont {Gondolf}, \citenamefont {B\"otzel},\ and\ \citenamefont {Eremin}}]{lechermann_electronic_2023}%
  \BibitemOpen
  \bibfield  {author} {\bibinfo {author} {\bibfnamefont {F.}~\bibnamefont {Lechermann}}, \bibinfo {author} {\bibfnamefont {J.}~\bibnamefont {Gondolf}}, \bibinfo {author} {\bibfnamefont {S.}~\bibnamefont {B\"otzel}}, \ and\ \bibinfo {author} {\bibfnamefont {I.~M.}\ \bibnamefont {Eremin}},\ }\href {\doibase 10.1103/PhysRevB.108.L201121} {\bibfield  {journal} {\bibinfo  {journal} {Phys. Rev. B}\ }\textbf {\bibinfo {volume} {108}},\ \bibinfo {pages} {L201121} (\bibinfo {year} {2023})}\BibitemShut {NoStop}%
\bibitem [{\citenamefont {Heier}\ \emph {et~al.}(2024)\citenamefont {Heier}, \citenamefont {Park},\ and\ \citenamefont {Savrasov}}]{heier_competing_2024}%
  \BibitemOpen
  \bibfield  {author} {\bibinfo {author} {\bibfnamefont {G.}~\bibnamefont {Heier}}, \bibinfo {author} {\bibfnamefont {K.}~\bibnamefont {Park}}, \ and\ \bibinfo {author} {\bibfnamefont {S.~Y.}\ \bibnamefont {Savrasov}},\ }\href {\doibase 10.1103/PhysRevB.109.104508} {\bibfield  {journal} {\bibinfo  {journal} {Phys. Rev. B}\ }\textbf {\bibinfo {volume} {109}},\ \bibinfo {pages} {104508} (\bibinfo {year} {2024})}\BibitemShut {NoStop}%
\bibitem [{\citenamefont {Jiang}\ \emph {et~al.}(2024)\citenamefont {Jiang}, \citenamefont {Wang},\ and\ \citenamefont {Zhang}}]{jiang_high_2024}%
  \BibitemOpen
  \bibfield  {author} {\bibinfo {author} {\bibfnamefont {K.}~\bibnamefont {Jiang}}, \bibinfo {author} {\bibfnamefont {Z.}~\bibnamefont {Wang}}, \ and\ \bibinfo {author} {\bibfnamefont {F.-C.}\ \bibnamefont {Zhang}},\ }\href {\doibase 10.1088/0256-307x/41/1/017402} {\bibfield  {journal} {\bibinfo  {journal} {Chin. Phys. Lett.}\ }\textbf {\bibinfo {volume} {41}},\ \bibinfo {pages} {017402} (\bibinfo {year} {2024})}\BibitemShut {NoStop}%
\bibitem [{\citenamefont {Zhang}\ \emph {et~al.}(2020)\citenamefont {Zhang}, \citenamefont {Phelan}, \citenamefont {Botana}, \citenamefont {Chen}, \citenamefont {Zheng}, \citenamefont {Krogstad}, \citenamefont {Wang}, \citenamefont {Qiu}, \citenamefont {Rodriguez-Rivera}, \citenamefont {Osborn}, \citenamefont {Rosenkranz}, \citenamefont {Norman},\ and\ \citenamefont {Mitchell}}]{zhang_intertwined_2020}%
  \BibitemOpen
  \bibfield  {author} {\bibinfo {author} {\bibfnamefont {J.}~\bibnamefont {Zhang}}, \bibinfo {author} {\bibfnamefont {D.}~\bibnamefont {Phelan}}, \bibinfo {author} {\bibfnamefont {A.~S.}\ \bibnamefont {Botana}}, \bibinfo {author} {\bibfnamefont {Y.-S.}\ \bibnamefont {Chen}}, \bibinfo {author} {\bibfnamefont {H.}~\bibnamefont {Zheng}}, \bibinfo {author} {\bibfnamefont {M.}~\bibnamefont {Krogstad}}, \bibinfo {author} {\bibfnamefont {S.~G.}\ \bibnamefont {Wang}}, \bibinfo {author} {\bibfnamefont {Y.}~\bibnamefont {Qiu}}, \bibinfo {author} {\bibfnamefont {J.~A.}\ \bibnamefont {Rodriguez-Rivera}}, \bibinfo {author} {\bibfnamefont {R.}~\bibnamefont {Osborn}}, \bibinfo {author} {\bibfnamefont {S.}~\bibnamefont {Rosenkranz}}, \bibinfo {author} {\bibfnamefont {M.~R.}\ \bibnamefont {Norman}}, \ and\ \bibinfo {author} {\bibfnamefont {J.~F.}\ \bibnamefont {Mitchell}},\ }\href {\doibase 10.1038/s41467-020-19836-0} {\bibfield  {journal} {\bibinfo  {journal} {Nat. Commun.}\ }\textbf {\bibinfo {volume} {11}},\ \bibinfo
  {pages} {6003} (\bibinfo {year} {2020})}\BibitemShut {NoStop}%
\bibitem [{\citenamefont {Ren}\ \emph {et~al.}(2025)\citenamefont {Ren}, \citenamefont {Sutarto}, \citenamefont {Wu}, \citenamefont {Zhang}, \citenamefont {Huang}, \citenamefont {Xiang}, \citenamefont {Hu}, \citenamefont {Comin}, \citenamefont {Zhou},\ and\ \citenamefont {Zhu}}]{ren_resolving_2025}%
  \BibitemOpen
  \bibfield  {author} {\bibinfo {author} {\bibfnamefont {X.}~\bibnamefont {Ren}}, \bibinfo {author} {\bibfnamefont {R.}~\bibnamefont {Sutarto}}, \bibinfo {author} {\bibfnamefont {X.}~\bibnamefont {Wu}}, \bibinfo {author} {\bibfnamefont {J.}~\bibnamefont {Zhang}}, \bibinfo {author} {\bibfnamefont {H.}~\bibnamefont {Huang}}, \bibinfo {author} {\bibfnamefont {T.}~\bibnamefont {Xiang}}, \bibinfo {author} {\bibfnamefont {J.}~\bibnamefont {Hu}}, \bibinfo {author} {\bibfnamefont {R.}~\bibnamefont {Comin}}, \bibinfo {author} {\bibfnamefont {X.}~\bibnamefont {Zhou}}, \ and\ \bibinfo {author} {\bibfnamefont {Z.}~\bibnamefont {Zhu}},\ }\href {\doibase 10.1038/s42005-025-01971-z} {\bibfield  {journal} {\bibinfo  {journal} {Commun. Phys.}\ }\textbf {\bibinfo {volume} {8}},\ \bibinfo {pages} {52} (\bibinfo {year} {2025})}\BibitemShut {NoStop}%
\bibitem [{\citenamefont {Gupta}\ \emph {et~al.}(2025)\citenamefont {Gupta}, \citenamefont {Gong}, \citenamefont {Wu}, \citenamefont {Kang}, \citenamefont {Parzyck}, \citenamefont {Gregory}, \citenamefont {Costa}, \citenamefont {Sutarto}, \citenamefont {Sarker}, \citenamefont {Singer}, \citenamefont {Schlom}, \citenamefont {Shen},\ and\ \citenamefont {Hawthorn}}]{gupta_anisotropic_2025}%
  \BibitemOpen
  \bibfield  {author} {\bibinfo {author} {\bibfnamefont {N.~K.}\ \bibnamefont {Gupta}}, \bibinfo {author} {\bibfnamefont {R.}~\bibnamefont {Gong}}, \bibinfo {author} {\bibfnamefont {Y.}~\bibnamefont {Wu}}, \bibinfo {author} {\bibfnamefont {M.}~\bibnamefont {Kang}}, \bibinfo {author} {\bibfnamefont {C.~T.}\ \bibnamefont {Parzyck}}, \bibinfo {author} {\bibfnamefont {B.~Z.}\ \bibnamefont {Gregory}}, \bibinfo {author} {\bibfnamefont {N.}~\bibnamefont {Costa}}, \bibinfo {author} {\bibfnamefont {R.}~\bibnamefont {Sutarto}}, \bibinfo {author} {\bibfnamefont {S.}~\bibnamefont {Sarker}}, \bibinfo {author} {\bibfnamefont {A.}~\bibnamefont {Singer}}, \bibinfo {author} {\bibfnamefont {D.~G.}\ \bibnamefont {Schlom}}, \bibinfo {author} {\bibfnamefont {K.~M.}\ \bibnamefont {Shen}}, \ and\ \bibinfo {author} {\bibfnamefont {D.~G.}\ \bibnamefont {Hawthorn}},\ }\href {\doibase 10.1038/s41467-025-61653-w} {\bibfield  {journal} {\bibinfo  {journal} {Nat. Commun.}\ }\textbf {\bibinfo {volume} {16}},\ \bibinfo {pages} {6560}
  (\bibinfo {year} {2025})}\BibitemShut {NoStop}%
\bibitem [{\citenamefont {Hepting}\ \emph {et~al.}(2020)\citenamefont {Hepting}, \citenamefont {Li}, \citenamefont {Jia}, \citenamefont {Lu}, \citenamefont {Paris}, \citenamefont {Tseng}, \citenamefont {Feng}, \citenamefont {Osada}, \citenamefont {Been}, \citenamefont {Hikita}, \citenamefont {Chuang}, \citenamefont {Hussain}, \citenamefont {Zhou}, \citenamefont {Nag}, \citenamefont {Garcia-Fernandez}, \citenamefont {Rossi}, \citenamefont {Huang}, \citenamefont {Huang}, \citenamefont {Shen}, \citenamefont {Schmitt}, \citenamefont {Hwang}, \citenamefont {Moritz}, \citenamefont {Zaanen}, \citenamefont {Devereaux},\ and\ \citenamefont {Lee}}]{hepting_electronic_2020}%
  \BibitemOpen
  \bibfield  {author} {\bibinfo {author} {\bibfnamefont {M.}~\bibnamefont {Hepting}}, \bibinfo {author} {\bibfnamefont {D.}~\bibnamefont {Li}}, \bibinfo {author} {\bibfnamefont {C.~J.}\ \bibnamefont {Jia}}, \bibinfo {author} {\bibfnamefont {H.}~\bibnamefont {Lu}}, \bibinfo {author} {\bibfnamefont {E.}~\bibnamefont {Paris}}, \bibinfo {author} {\bibfnamefont {Y.}~\bibnamefont {Tseng}}, \bibinfo {author} {\bibfnamefont {X.}~\bibnamefont {Feng}}, \bibinfo {author} {\bibfnamefont {M.}~\bibnamefont {Osada}}, \bibinfo {author} {\bibfnamefont {E.}~\bibnamefont {Been}}, \bibinfo {author} {\bibfnamefont {Y.}~\bibnamefont {Hikita}}, \bibinfo {author} {\bibfnamefont {Y.-D.}\ \bibnamefont {Chuang}}, \bibinfo {author} {\bibfnamefont {Z.}~\bibnamefont {Hussain}}, \bibinfo {author} {\bibfnamefont {K.~J.}\ \bibnamefont {Zhou}}, \bibinfo {author} {\bibfnamefont {A.}~\bibnamefont {Nag}}, \bibinfo {author} {\bibfnamefont {M.}~\bibnamefont {Garcia-Fernandez}}, \bibinfo {author} {\bibfnamefont {M.}~\bibnamefont {Rossi}}, \bibinfo
  {author} {\bibfnamefont {H.~Y.}\ \bibnamefont {Huang}}, \bibinfo {author} {\bibfnamefont {D.~J.}\ \bibnamefont {Huang}}, \bibinfo {author} {\bibfnamefont {Z.~X.}\ \bibnamefont {Shen}}, \bibinfo {author} {\bibfnamefont {T.}~\bibnamefont {Schmitt}}, \bibinfo {author} {\bibfnamefont {H.~Y.}\ \bibnamefont {Hwang}}, \bibinfo {author} {\bibfnamefont {B.}~\bibnamefont {Moritz}}, \bibinfo {author} {\bibfnamefont {J.}~\bibnamefont {Zaanen}}, \bibinfo {author} {\bibfnamefont {T.~P.}\ \bibnamefont {Devereaux}}, \ and\ \bibinfo {author} {\bibfnamefont {W.~S.}\ \bibnamefont {Lee}},\ }\href {\doibase 10.1038/s41563-019-0585-z} {\bibfield  {journal} {\bibinfo  {journal} {Nat. Mater.}\ }\textbf {\bibinfo {volume} {19}},\ \bibinfo {pages} {381} (\bibinfo {year} {2020})}\BibitemShut {NoStop}%
\bibitem [{\citenamefont {TenHuisen}\ \emph {et~al.}(2025)\citenamefont {TenHuisen}, \citenamefont {Pan}, \citenamefont {Song}, \citenamefont {Baykusheva}, \citenamefont {Ferenc~Segedin}, \citenamefont {Goodge}, \citenamefont {Paik}, \citenamefont {Pelliciari}, \citenamefont {Bisogni}, \citenamefont {Gu}, \citenamefont {Agrestini}, \citenamefont {Nag}, \citenamefont {Garc\'{\i}a-Fern\'andez}, \citenamefont {Zhou}, \citenamefont {Kourkoutis}, \citenamefont {Brooks}, \citenamefont {Mundy}, \citenamefont {Dean},\ and\ \citenamefont {Mitrano}}]{tenhuisen_magnetic_2025}%
  \BibitemOpen
  \bibfield  {author} {\bibinfo {author} {\bibfnamefont {S.~F.~R.}\ \bibnamefont {TenHuisen}}, \bibinfo {author} {\bibfnamefont {G.~A.}\ \bibnamefont {Pan}}, \bibinfo {author} {\bibfnamefont {Q.}~\bibnamefont {Song}}, \bibinfo {author} {\bibfnamefont {D.~R.}\ \bibnamefont {Baykusheva}}, \bibinfo {author} {\bibfnamefont {D.}~\bibnamefont {Ferenc~Segedin}}, \bibinfo {author} {\bibfnamefont {B.~H.}\ \bibnamefont {Goodge}}, \bibinfo {author} {\bibfnamefont {H.}~\bibnamefont {Paik}}, \bibinfo {author} {\bibfnamefont {J.}~\bibnamefont {Pelliciari}}, \bibinfo {author} {\bibfnamefont {V.}~\bibnamefont {Bisogni}}, \bibinfo {author} {\bibfnamefont {Y.}~\bibnamefont {Gu}}, \bibinfo {author} {\bibfnamefont {S.}~\bibnamefont {Agrestini}}, \bibinfo {author} {\bibfnamefont {A.}~\bibnamefont {Nag}}, \bibinfo {author} {\bibfnamefont {M.}~\bibnamefont {Garc\'{\i}a-Fern\'andez}}, \bibinfo {author} {\bibfnamefont {K.-J.}\ \bibnamefont {Zhou}}, \bibinfo {author} {\bibfnamefont {L.~F.}\ \bibnamefont {Kourkoutis}}, \bibinfo
  {author} {\bibfnamefont {C.~M.}\ \bibnamefont {Brooks}}, \bibinfo {author} {\bibfnamefont {J.~A.}\ \bibnamefont {Mundy}}, \bibinfo {author} {\bibfnamefont {M.~P.~M.}\ \bibnamefont {Dean}}, \ and\ \bibinfo {author} {\bibfnamefont {M.}~\bibnamefont {Mitrano}},\ }\href {\doibase 10.1103/PhysRevB.111.165145} {\bibfield  {journal} {\bibinfo  {journal} {Phys. Rev. B}\ }\textbf {\bibinfo {volume} {111}},\ \bibinfo {pages} {165145} (\bibinfo {year} {2025})}\BibitemShut {NoStop}%
\bibitem [{\citenamefont {Zhang}\ \emph {et~al.}(2025)\citenamefont {Zhang}, \citenamefont {Zhang}, \citenamefont {Dong}, \citenamefont {Li}, \citenamefont {Xiao}, \citenamefont {Huo}, \citenamefont {Huang}, \citenamefont {Huang}, \citenamefont {Wang}, \citenamefont {Lu}, \citenamefont {Chen}, \citenamefont {Wang},\ and\ \citenamefont {Peng}}]{zhang_distinct_2025}%
  \BibitemOpen
  \bibfield  {author} {\bibinfo {author} {\bibfnamefont {S.}~\bibnamefont {Zhang}}, \bibinfo {author} {\bibfnamefont {H.}~\bibnamefont {Zhang}}, \bibinfo {author} {\bibfnamefont {Z.}~\bibnamefont {Dong}}, \bibinfo {author} {\bibfnamefont {J.}~\bibnamefont {Li}}, \bibinfo {author} {\bibfnamefont {Q.}~\bibnamefont {Xiao}}, \bibinfo {author} {\bibfnamefont {M.}~\bibnamefont {Huo}}, \bibinfo {author} {\bibfnamefont {H.-Y.}\ \bibnamefont {Huang}}, \bibinfo {author} {\bibfnamefont {D.-J.}\ \bibnamefont {Huang}}, \bibinfo {author} {\bibfnamefont {Y.}~\bibnamefont {Wang}}, \bibinfo {author} {\bibfnamefont {Y.}~\bibnamefont {Lu}}, \bibinfo {author} {\bibfnamefont {Z.}~\bibnamefont {Chen}}, \bibinfo {author} {\bibfnamefont {M.}~\bibnamefont {Wang}}, \ and\ \bibinfo {author} {\bibfnamefont {Y.}~\bibnamefont {Peng}},\ }\href {https://arxiv.org/abs/2509.20727} {\bibfield  {journal} {\bibinfo  {journal} {arXiv:2509.20727}\ } (\bibinfo {year} {2025})}\BibitemShut {NoStop}%
\bibitem [{\citenamefont {Lamsal}\ and\ \citenamefont {Montfrooij}(2016)}]{lamsal_extracting_2016}%
  \BibitemOpen
  \bibfield  {author} {\bibinfo {author} {\bibfnamefont {J.}~\bibnamefont {Lamsal}}\ and\ \bibinfo {author} {\bibfnamefont {W.}~\bibnamefont {Montfrooij}},\ }\href {\doibase 10.1103/PhysRevB.93.214513} {\bibfield  {journal} {\bibinfo  {journal} {Phys. Rev. B}\ }\textbf {\bibinfo {volume} {93}},\ \bibinfo {pages} {214513} (\bibinfo {year} {2016})}\BibitemShut {NoStop}%
\bibitem [{\citenamefont {Yan}\ \emph {et~al.}(2025)\citenamefont {Yan}, \citenamefont {Chan}, \citenamefont {Hong}, \citenamefont {Chow}, \citenamefont {Luo}, \citenamefont {Li}, \citenamefont {Wang}, \citenamefont {Wu}, \citenamefont {Biało}, \citenamefont {Fitriyah}, \citenamefont {Prakash}, \citenamefont {Gao}, \citenamefont {Yip}, \citenamefont {Gao}, \citenamefont {Ren}, \citenamefont {Choi}, \citenamefont {Channagowdra}, \citenamefont {Okamoto}, \citenamefont {Zhou}, \citenamefont {Zhu}, \citenamefont {Si}, \citenamefont {Garcia-Fernandez}, \citenamefont {Zhou}, \citenamefont {Huang}, \citenamefont {Huang}, \citenamefont {Chang}, \citenamefont {Ariando},\ and\ \citenamefont {Wang}}]{yan_persistent_2025}%
  \BibitemOpen
  \bibfield  {author} {\bibinfo {author} {\bibfnamefont {Y.}~\bibnamefont {Yan}}, \bibinfo {author} {\bibfnamefont {Y.}~\bibnamefont {Chan}}, \bibinfo {author} {\bibfnamefont {X.}~\bibnamefont {Hong}}, \bibinfo {author} {\bibfnamefont {S.~L.~E.}\ \bibnamefont {Chow}}, \bibinfo {author} {\bibfnamefont {Z.}~\bibnamefont {Luo}}, \bibinfo {author} {\bibfnamefont {Y.}~\bibnamefont {Li}}, \bibinfo {author} {\bibfnamefont {T.}~\bibnamefont {Wang}}, \bibinfo {author} {\bibfnamefont {Y.}~\bibnamefont {Wu}}, \bibinfo {author} {\bibfnamefont {I.}~\bibnamefont {Biało}}, \bibinfo {author} {\bibfnamefont {N.}~\bibnamefont {Fitriyah}}, \bibinfo {author} {\bibfnamefont {S.}~\bibnamefont {Prakash}}, \bibinfo {author} {\bibfnamefont {X.}~\bibnamefont {Gao}}, \bibinfo {author} {\bibfnamefont {K.~Y.}\ \bibnamefont {Yip}}, \bibinfo {author} {\bibfnamefont {Q.}~\bibnamefont {Gao}}, \bibinfo {author} {\bibfnamefont {X.}~\bibnamefont {Ren}}, \bibinfo {author} {\bibfnamefont {J.}~\bibnamefont {Choi}}, \bibinfo {author}
  {\bibfnamefont {G.}~\bibnamefont {Channagowdra}}, \bibinfo {author} {\bibfnamefont {J.}~\bibnamefont {Okamoto}}, \bibinfo {author} {\bibfnamefont {X.}~\bibnamefont {Zhou}}, \bibinfo {author} {\bibfnamefont {Z.}~\bibnamefont {Zhu}}, \bibinfo {author} {\bibfnamefont {L.}~\bibnamefont {Si}}, \bibinfo {author} {\bibfnamefont {M.}~\bibnamefont {Garcia-Fernandez}}, \bibinfo {author} {\bibfnamefont {K.-J.}\ \bibnamefont {Zhou}}, \bibinfo {author} {\bibfnamefont {H.-Y.}\ \bibnamefont {Huang}}, \bibinfo {author} {\bibfnamefont {D.-J.}\ \bibnamefont {Huang}}, \bibinfo {author} {\bibfnamefont {J.}~\bibnamefont {Chang}}, \bibinfo {author} {\bibfnamefont {A.}~\bibnamefont {Ariando}}, \ and\ \bibinfo {author} {\bibfnamefont {Q.}~\bibnamefont {Wang}},\ }\href {https://arxiv.org/abs/2507.18373} {\bibfield  {journal} {\bibinfo  {journal} {arXiv:2507.18373}\ } (\bibinfo {year} {2025})}\BibitemShut {NoStop}%
\bibitem [{\citenamefont {Shi}\ \emph {et~al.}(2025)\citenamefont {Shi}, \citenamefont {Li}, \citenamefont {Wang}, \citenamefont {Peng}, \citenamefont {Yang}, \citenamefont {Li}, \citenamefont {Fan}, \citenamefont {Jiang}, \citenamefont {He}, \citenamefont {Zeng}, \citenamefont {Song}, \citenamefont {Ge}, \citenamefont {Xiang}, \citenamefont {Wang}, \citenamefont {Ying}, \citenamefont {Wu},\ and\ \citenamefont {Chen}}]{shi_absence_2025}%
  \BibitemOpen
  \bibfield  {author} {\bibinfo {author} {\bibfnamefont {M.}~\bibnamefont {Shi}}, \bibinfo {author} {\bibfnamefont {Y.}~\bibnamefont {Li}}, \bibinfo {author} {\bibfnamefont {Y.}~\bibnamefont {Wang}}, \bibinfo {author} {\bibfnamefont {D.}~\bibnamefont {Peng}}, \bibinfo {author} {\bibfnamefont {S.}~\bibnamefont {Yang}}, \bibinfo {author} {\bibfnamefont {H.}~\bibnamefont {Li}}, \bibinfo {author} {\bibfnamefont {K.}~\bibnamefont {Fan}}, \bibinfo {author} {\bibfnamefont {K.}~\bibnamefont {Jiang}}, \bibinfo {author} {\bibfnamefont {J.}~\bibnamefont {He}}, \bibinfo {author} {\bibfnamefont {Q.}~\bibnamefont {Zeng}}, \bibinfo {author} {\bibfnamefont {D.}~\bibnamefont {Song}}, \bibinfo {author} {\bibfnamefont {B.}~\bibnamefont {Ge}}, \bibinfo {author} {\bibfnamefont {Z.}~\bibnamefont {Xiang}}, \bibinfo {author} {\bibfnamefont {Z.}~\bibnamefont {Wang}}, \bibinfo {author} {\bibfnamefont {J.}~\bibnamefont {Ying}}, \bibinfo {author} {\bibfnamefont {T.}~\bibnamefont {Wu}}, \ and\ \bibinfo {author} {\bibfnamefont
  {X.}~\bibnamefont {Chen}},\ }\href {\doibase 10.1038/s41467-025-57264-0} {\bibfield  {journal} {\bibinfo  {journal} {Nat. Commun.}\ }\textbf {\bibinfo {volume} {16}},\ \bibinfo {pages} {2887} (\bibinfo {year} {2025})}\BibitemShut {NoStop}%
\bibitem [{\citenamefont {Pei}\ \emph {et~al.}(2025)\citenamefont {Pei}, \citenamefont {Shen}, \citenamefont {Peng}, \citenamefont {Zhang}, \citenamefont {Zhao}, \citenamefont {Xing}, \citenamefont {Wang}, \citenamefont {Wu}, \citenamefont {Wang}, \citenamefont {Zhao}, \citenamefont {Xing}, \citenamefont {Chen}, \citenamefont {Zhao}, \citenamefont {Yang}, \citenamefont {Liu}, \citenamefont {Shi}, \citenamefont {Guo}, \citenamefont {Zeng}, \citenamefont {Zhang},\ and\ \citenamefont {Qi}}]{pei_weakly_2025}%
  \BibitemOpen
  \bibfield  {author} {\bibinfo {author} {\bibfnamefont {C.}~\bibnamefont {Pei}}, \bibinfo {author} {\bibfnamefont {Y.}~\bibnamefont {Shen}}, \bibinfo {author} {\bibfnamefont {D.}~\bibnamefont {Peng}}, \bibinfo {author} {\bibfnamefont {M.}~\bibnamefont {Zhang}}, \bibinfo {author} {\bibfnamefont {Y.}~\bibnamefont {Zhao}}, \bibinfo {author} {\bibfnamefont {X.}~\bibnamefont {Xing}}, \bibinfo {author} {\bibfnamefont {Q.}~\bibnamefont {Wang}}, \bibinfo {author} {\bibfnamefont {J.}~\bibnamefont {Wu}}, \bibinfo {author} {\bibfnamefont {J.}~\bibnamefont {Wang}}, \bibinfo {author} {\bibfnamefont {L.}~\bibnamefont {Zhao}}, \bibinfo {author} {\bibfnamefont {Z.}~\bibnamefont {Xing}}, \bibinfo {author} {\bibfnamefont {Y.}~\bibnamefont {Chen}}, \bibinfo {author} {\bibfnamefont {J.}~\bibnamefont {Zhao}}, \bibinfo {author} {\bibfnamefont {W.}~\bibnamefont {Yang}}, \bibinfo {author} {\bibfnamefont {X.}~\bibnamefont {Liu}}, \bibinfo {author} {\bibfnamefont {Z.}~\bibnamefont {Shi}}, \bibinfo {author} {\bibfnamefont
  {H.}~\bibnamefont {Guo}}, \bibinfo {author} {\bibfnamefont {Q.}~\bibnamefont {Zeng}}, \bibinfo {author} {\bibfnamefont {G.-M.}\ \bibnamefont {Zhang}}, \ and\ \bibinfo {author} {\bibfnamefont {Y.}~\bibnamefont {Qi}},\ }\href {\doibase 10.1021/jacs.5c17977} {\bibfield  {journal} {\bibinfo  {journal} {J. Am. Chem. Soc.}\ }\textbf {\bibinfo {volume} {148}},\ \bibinfo {pages} {1388} (\bibinfo {year} {2025})}\BibitemShut {NoStop}%
\bibitem [{\citenamefont {Yang}\ \emph {et~al.}(2026)\citenamefont {Yang}, \citenamefont {Zhan}, \citenamefont {Miao}, \citenamefont {Huo}, \citenamefont {Xu}, \citenamefont {Li}, \citenamefont {Xie}, \citenamefont {Liang}, \citenamefont {Cai}, \citenamefont {Chen}, \citenamefont {Zhu}, \citenamefont {Xu}, \citenamefont {Zhang}, \citenamefont {Zhang}, \citenamefont {Yang}, \citenamefont {Wang}, \citenamefont {Peng}, \citenamefont {Mao}, \citenamefont {Li}, \citenamefont {Zhu}, \citenamefont {Liu}, \citenamefont {Xu}, \citenamefont {Hu}, \citenamefont {Wu}, \citenamefont {Wang}, \citenamefont {Zhao},\ and\ \citenamefont {Zhou}}]{yang_electronic_2026}%
  \BibitemOpen
  \bibfield  {author} {\bibinfo {author} {\bibfnamefont {J.}~\bibnamefont {Yang}}, \bibinfo {author} {\bibfnamefont {J.}~\bibnamefont {Zhan}}, \bibinfo {author} {\bibfnamefont {T.}~\bibnamefont {Miao}}, \bibinfo {author} {\bibfnamefont {M.}~\bibnamefont {Huo}}, \bibinfo {author} {\bibfnamefont {Q.}~\bibnamefont {Xu}}, \bibinfo {author} {\bibfnamefont {Y.}~\bibnamefont {Li}}, \bibinfo {author} {\bibfnamefont {Y.}~\bibnamefont {Xie}}, \bibinfo {author} {\bibfnamefont {B.}~\bibnamefont {Liang}}, \bibinfo {author} {\bibfnamefont {N.}~\bibnamefont {Cai}}, \bibinfo {author} {\bibfnamefont {H.}~\bibnamefont {Chen}}, \bibinfo {author} {\bibfnamefont {W.}~\bibnamefont {Zhu}}, \bibinfo {author} {\bibfnamefont {M.}~\bibnamefont {Xu}}, \bibinfo {author} {\bibfnamefont {S.}~\bibnamefont {Zhang}}, \bibinfo {author} {\bibfnamefont {F.}~\bibnamefont {Zhang}}, \bibinfo {author} {\bibfnamefont {F.}~\bibnamefont {Yang}}, \bibinfo {author} {\bibfnamefont {Z.}~\bibnamefont {Wang}}, \bibinfo {author} {\bibfnamefont
  {Q.}~\bibnamefont {Peng}}, \bibinfo {author} {\bibfnamefont {H.}~\bibnamefont {Mao}}, \bibinfo {author} {\bibfnamefont {X.}~\bibnamefont {Li}}, \bibinfo {author} {\bibfnamefont {Z.}~\bibnamefont {Zhu}}, \bibinfo {author} {\bibfnamefont {G.}~\bibnamefont {Liu}}, \bibinfo {author} {\bibfnamefont {Z.}~\bibnamefont {Xu}}, \bibinfo {author} {\bibfnamefont {J.}~\bibnamefont {Hu}}, \bibinfo {author} {\bibfnamefont {X.}~\bibnamefont {Wu}}, \bibinfo {author} {\bibfnamefont {M.}~\bibnamefont {Wang}}, \bibinfo {author} {\bibfnamefont {L.}~\bibnamefont {Zhao}}, \ and\ \bibinfo {author} {\bibfnamefont {X.~J.}\ \bibnamefont {Zhou}},\ }\href {https://arxiv.org/abs/2601.22608} {\bibfield  {journal} {\bibinfo  {journal} {arXiv:2601.22608}\ } (\bibinfo {year} {2026})}\BibitemShut {NoStop}%
\bibitem [{\citenamefont {Jiang}\ \emph {et~al.}(2026)\citenamefont {Jiang}, \citenamefont {Zhang}, \citenamefont {Wang}, \citenamefont {Liu}, \citenamefont {Liu}, \citenamefont {Zhang}, \citenamefont {Zhang}, \citenamefont {Jing}, \citenamefont {Huang}, \citenamefont {Jiang}, \citenamefont {Ye}, \citenamefont {Jiang}, \citenamefont {Zhao}, \citenamefont {Shen},\ and\ \citenamefont {Feng}}]{jiang_direct_2026}%
  \BibitemOpen
  \bibfield  {author} {\bibinfo {author} {\bibfnamefont {Z.}~\bibnamefont {Jiang}}, \bibinfo {author} {\bibfnamefont {E.}~\bibnamefont {Zhang}}, \bibinfo {author} {\bibfnamefont {Y.}~\bibnamefont {Wang}}, \bibinfo {author} {\bibfnamefont {Z.}~\bibnamefont {Liu}}, \bibinfo {author} {\bibfnamefont {J.}~\bibnamefont {Liu}}, \bibinfo {author} {\bibfnamefont {R.}~\bibnamefont {Zhang}}, \bibinfo {author} {\bibfnamefont {X.}~\bibnamefont {Zhang}}, \bibinfo {author} {\bibfnamefont {W.}~\bibnamefont {Jing}}, \bibinfo {author} {\bibfnamefont {Y.}~\bibnamefont {Huang}}, \bibinfo {author} {\bibfnamefont {Q.}~\bibnamefont {Jiang}}, \bibinfo {author} {\bibfnamefont {M.}~\bibnamefont {Ye}}, \bibinfo {author} {\bibfnamefont {K.}~\bibnamefont {Jiang}}, \bibinfo {author} {\bibfnamefont {J.}~\bibnamefont {Zhao}}, \bibinfo {author} {\bibfnamefont {D.}~\bibnamefont {Shen}}, \ and\ \bibinfo {author} {\bibfnamefont {D.}~\bibnamefont {Feng}},\ }\href {https://arxiv.org/abs/2602.02127} {\bibfield  {journal} {\bibinfo  {journal}
  {arXiv:2602.02127}\ } (\bibinfo {year} {2026})}\BibitemShut {NoStop}%
\bibitem [{\citenamefont {Li}\ \emph {et~al.}(2026)\citenamefont {Li}, \citenamefont {Zhang}, \citenamefont {Du}, \citenamefont {Pei}, \citenamefont {Liu}, \citenamefont {Chen}, \citenamefont {Zhao}, \citenamefont {Zhai}, \citenamefont {Hu}, \citenamefont {Zhang}, \citenamefont {Shao}, \citenamefont {Mao}, \citenamefont {Cao}, \citenamefont {Zhao}, \citenamefont {Li}, \citenamefont {Shen}, \citenamefont {Huang}, \citenamefont {Hashimoto}, \citenamefont {Lu}, \citenamefont {Liu}, \citenamefont {Chen}, \citenamefont {Guo}, \citenamefont {Wang}, \citenamefont {Qi},\ and\ \citenamefont {Yang}}]{li_orbital-selective_2026}%
  \BibitemOpen
  \bibfield  {author} {\bibinfo {author} {\bibfnamefont {Y.}~\bibnamefont {Li}}, \bibinfo {author} {\bibfnamefont {M.}~\bibnamefont {Zhang}}, \bibinfo {author} {\bibfnamefont {X.}~\bibnamefont {Du}}, \bibinfo {author} {\bibfnamefont {C.}~\bibnamefont {Pei}}, \bibinfo {author} {\bibfnamefont {J.}~\bibnamefont {Liu}}, \bibinfo {author} {\bibfnamefont {H.}~\bibnamefont {Chen}}, \bibinfo {author} {\bibfnamefont {W.}~\bibnamefont {Zhao}}, \bibinfo {author} {\bibfnamefont {K.}~\bibnamefont {Zhai}}, \bibinfo {author} {\bibfnamefont {Y.}~\bibnamefont {Hu}}, \bibinfo {author} {\bibfnamefont {S.}~\bibnamefont {Zhang}}, \bibinfo {author} {\bibfnamefont {J.}~\bibnamefont {Shao}}, \bibinfo {author} {\bibfnamefont {M.}~\bibnamefont {Mao}}, \bibinfo {author} {\bibfnamefont {Y.}~\bibnamefont {Cao}}, \bibinfo {author} {\bibfnamefont {J.}~\bibnamefont {Zhao}}, \bibinfo {author} {\bibfnamefont {Z.}~\bibnamefont {Li}}, \bibinfo {author} {\bibfnamefont {D.}~\bibnamefont {Shen}}, \bibinfo {author} {\bibfnamefont {Y.}~\bibnamefont
  {Huang}}, \bibinfo {author} {\bibfnamefont {M.}~\bibnamefont {Hashimoto}}, \bibinfo {author} {\bibfnamefont {D.}~\bibnamefont {Lu}}, \bibinfo {author} {\bibfnamefont {Z.}~\bibnamefont {Liu}}, \bibinfo {author} {\bibfnamefont {Y.}~\bibnamefont {Chen}}, \bibinfo {author} {\bibfnamefont {H.}~\bibnamefont {Guo}}, \bibinfo {author} {\bibfnamefont {Y.}~\bibnamefont {Wang}}, \bibinfo {author} {\bibfnamefont {Y.}~\bibnamefont {Qi}}, \ and\ \bibinfo {author} {\bibfnamefont {L.}~\bibnamefont {Yang}},\ }\href {https://arxiv.org/abs/2602.03658} {\bibfield  {journal} {\bibinfo  {journal} {arXiv:2602.03658}\ } (\bibinfo {year} {2026})}\BibitemShut {NoStop}%
\bibitem [{\citenamefont {Chen}\ \emph {et~al.}(2026)\citenamefont {Chen}, \citenamefont {Li}, \citenamefont {Xie}, \citenamefont {Hu}, \citenamefont {Chiu}, \citenamefont {Agrestini}, \citenamefont {Zhang}, \citenamefont {Lu}, \citenamefont {Wang}, \citenamefont {Garcia-Fernandez}, \citenamefont {Feng},\ and\ \citenamefont {Zhou}}]{chen_dissecting_2026}%
  \BibitemOpen
  \bibfield  {author} {\bibinfo {author} {\bibfnamefont {X.}~\bibnamefont {Chen}}, \bibinfo {author} {\bibfnamefont {Z.}~\bibnamefont {Li}}, \bibinfo {author} {\bibfnamefont {M.}~\bibnamefont {Xie}}, \bibinfo {author} {\bibfnamefont {D.}~\bibnamefont {Hu}}, \bibinfo {author} {\bibfnamefont {Y.-F.}\ \bibnamefont {Chiu}}, \bibinfo {author} {\bibfnamefont {S.}~\bibnamefont {Agrestini}}, \bibinfo {author} {\bibfnamefont {W.}~\bibnamefont {Zhang}}, \bibinfo {author} {\bibfnamefont {Y.}~\bibnamefont {Lu}}, \bibinfo {author} {\bibfnamefont {M.}~\bibnamefont {Wang}}, \bibinfo {author} {\bibfnamefont {M.}~\bibnamefont {Garcia-Fernandez}}, \bibinfo {author} {\bibfnamefont {D.}~\bibnamefont {Feng}}, \ and\ \bibinfo {author} {\bibfnamefont {K.-J.}\ \bibnamefont {Zhou}},\ }\href {http://arxiv.org/abs/2604.01902} {\bibfield  {journal} {\bibinfo  {journal} {arXiv:2604.01902}\ } (\bibinfo {year} {2026})}\BibitemShut {NoStop}%
\bibitem [{\citenamefont {Brookes}\ \emph {et~al.}(2018)\citenamefont {Brookes}, \citenamefont {Yakhou-Harris}, \citenamefont {Kummer}, \citenamefont {Fondacaro}, \citenamefont {Cezar}, \citenamefont {Betto}, \citenamefont {Velez-Fort}, \citenamefont {Amorese}, \citenamefont {Ghiringhelli}, \citenamefont {Braicovich}, \citenamefont {Barrett}, \citenamefont {Berruyer}, \citenamefont {Cianciosi}, \citenamefont {Eybert}, \citenamefont {Marion}, \citenamefont {van~der Linden},\ and\ \citenamefont {Zhang}}]{brookes_beamline_2018}%
  \BibitemOpen
  \bibfield  {author} {\bibinfo {author} {\bibfnamefont {N.~B.}\ \bibnamefont {Brookes}}, \bibinfo {author} {\bibfnamefont {F.}~\bibnamefont {Yakhou-Harris}}, \bibinfo {author} {\bibfnamefont {K.}~\bibnamefont {Kummer}}, \bibinfo {author} {\bibfnamefont {A.}~\bibnamefont {Fondacaro}}, \bibinfo {author} {\bibfnamefont {J.~C.}\ \bibnamefont {Cezar}}, \bibinfo {author} {\bibfnamefont {D.}~\bibnamefont {Betto}}, \bibinfo {author} {\bibfnamefont {E.}~\bibnamefont {Velez-Fort}}, \bibinfo {author} {\bibfnamefont {A.}~\bibnamefont {Amorese}}, \bibinfo {author} {\bibfnamefont {G.}~\bibnamefont {Ghiringhelli}}, \bibinfo {author} {\bibfnamefont {L.}~\bibnamefont {Braicovich}}, \bibinfo {author} {\bibfnamefont {R.}~\bibnamefont {Barrett}}, \bibinfo {author} {\bibfnamefont {G.}~\bibnamefont {Berruyer}}, \bibinfo {author} {\bibfnamefont {F.}~\bibnamefont {Cianciosi}}, \bibinfo {author} {\bibfnamefont {L.}~\bibnamefont {Eybert}}, \bibinfo {author} {\bibfnamefont {P.}~\bibnamefont {Marion}}, \bibinfo {author} {\bibfnamefont
  {P.}~\bibnamefont {van~der Linden}}, \ and\ \bibinfo {author} {\bibfnamefont {L.}~\bibnamefont {Zhang}},\ }\href {\doibase 10.1016/j.nima.2018.07.001} {\bibfield  {journal} {\bibinfo  {journal} {Nucl. Instrum. Methods Phys. Res. A.}\ }\textbf {\bibinfo {volume} {903}},\ \bibinfo {pages} {175} (\bibinfo {year} {2018})}\BibitemShut {NoStop}%
\bibitem [{\citenamefont {Wang}\ \emph {et~al.}(2021)\citenamefont {Wang}, \citenamefont {von Arx}, \citenamefont {Horio}, \citenamefont {Mukkattukavil}, \citenamefont {K{\"u}spert}, \citenamefont {Sassa}, \citenamefont {Schmitt}, \citenamefont {Nag}, \citenamefont {Pyon}, \citenamefont {Takayama}, \citenamefont {Takagi}, \citenamefont {Garcia-Fernandez}, \citenamefont {Zhou},\ and\ \citenamefont {Chang}}]{wang_charge_2021}%
  \BibitemOpen
  \bibfield  {author} {\bibinfo {author} {\bibfnamefont {Q.}~\bibnamefont {Wang}}, \bibinfo {author} {\bibfnamefont {K.}~\bibnamefont {von Arx}}, \bibinfo {author} {\bibfnamefont {M.}~\bibnamefont {Horio}}, \bibinfo {author} {\bibfnamefont {D.~J.}\ \bibnamefont {Mukkattukavil}}, \bibinfo {author} {\bibfnamefont {J.}~\bibnamefont {K{\"u}spert}}, \bibinfo {author} {\bibfnamefont {Y.}~\bibnamefont {Sassa}}, \bibinfo {author} {\bibfnamefont {T.}~\bibnamefont {Schmitt}}, \bibinfo {author} {\bibfnamefont {A.}~\bibnamefont {Nag}}, \bibinfo {author} {\bibfnamefont {S.}~\bibnamefont {Pyon}}, \bibinfo {author} {\bibfnamefont {T.}~\bibnamefont {Takayama}}, \bibinfo {author} {\bibfnamefont {H.}~\bibnamefont {Takagi}}, \bibinfo {author} {\bibfnamefont {M.}~\bibnamefont {Garcia-Fernandez}}, \bibinfo {author} {\bibfnamefont {K.-J.}\ \bibnamefont {Zhou}}, \ and\ \bibinfo {author} {\bibfnamefont {J.}~\bibnamefont {Chang}},\ }\href {\doibase 10.1126/sciadv.abg7394} {\bibfield  {journal} {\bibinfo  {journal} {Sci. Adv.}\
  }\textbf {\bibinfo {volume} {7}},\ \bibinfo {pages} {eabg7394} (\bibinfo {year} {2021})}\BibitemShut {NoStop}%
\bibitem [{\citenamefont {Arpaia}\ \emph {et~al.}(2023)\citenamefont {Arpaia}, \citenamefont {Martinelli}, \citenamefont {Sala}, \citenamefont {Caprara}, \citenamefont {Nag}, \citenamefont {Brookes}, \citenamefont {Camisa}, \citenamefont {Li}, \citenamefont {Gao}, \citenamefont {Zhou}, \citenamefont {Garcia-Fernandez}, \citenamefont {Zhou}, \citenamefont {Schierle}, \citenamefont {Bauch}, \citenamefont {Peng}, \citenamefont {Di~Castro}, \citenamefont {Grilli}, \citenamefont {Lombardi}, \citenamefont {Braicovich},\ and\ \citenamefont {Ghiringhelli}}]{arpaia_signature_2023}%
  \BibitemOpen
  \bibfield  {author} {\bibinfo {author} {\bibfnamefont {R.}~\bibnamefont {Arpaia}}, \bibinfo {author} {\bibfnamefont {L.}~\bibnamefont {Martinelli}}, \bibinfo {author} {\bibfnamefont {M.~M.}\ \bibnamefont {Sala}}, \bibinfo {author} {\bibfnamefont {S.}~\bibnamefont {Caprara}}, \bibinfo {author} {\bibfnamefont {A.}~\bibnamefont {Nag}}, \bibinfo {author} {\bibfnamefont {N.~B.}\ \bibnamefont {Brookes}}, \bibinfo {author} {\bibfnamefont {P.}~\bibnamefont {Camisa}}, \bibinfo {author} {\bibfnamefont {Q.}~\bibnamefont {Li}}, \bibinfo {author} {\bibfnamefont {Q.}~\bibnamefont {Gao}}, \bibinfo {author} {\bibfnamefont {X.}~\bibnamefont {Zhou}}, \bibinfo {author} {\bibfnamefont {M.}~\bibnamefont {Garcia-Fernandez}}, \bibinfo {author} {\bibfnamefont {K.-J.}\ \bibnamefont {Zhou}}, \bibinfo {author} {\bibfnamefont {E.}~\bibnamefont {Schierle}}, \bibinfo {author} {\bibfnamefont {T.}~\bibnamefont {Bauch}}, \bibinfo {author} {\bibfnamefont {Y.~Y.}\ \bibnamefont {Peng}}, \bibinfo {author} {\bibfnamefont {C.}~\bibnamefont
  {Di~Castro}}, \bibinfo {author} {\bibfnamefont {M.}~\bibnamefont {Grilli}}, \bibinfo {author} {\bibfnamefont {F.}~\bibnamefont {Lombardi}}, \bibinfo {author} {\bibfnamefont {L.}~\bibnamefont {Braicovich}}, \ and\ \bibinfo {author} {\bibfnamefont {G.}~\bibnamefont {Ghiringhelli}},\ }\href {\doibase 10.1038/s41467-023-42961-5} {\bibfield  {journal} {\bibinfo  {journal} {Nat. Commun.}\ }\textbf {\bibinfo {volume} {14}},\ \bibinfo {pages} {7198} (\bibinfo {year} {2023})}\BibitemShut {NoStop}%
\bibitem [{\citenamefont {Singh}\ \emph {et~al.}(2021)\citenamefont {Singh}, \citenamefont {Huang}, \citenamefont {Chu}, \citenamefont {Hua}, \citenamefont {Lin}, \citenamefont {Fung}, \citenamefont {Shiu}, \citenamefont {Chang}, \citenamefont {Li}, \citenamefont {Okamoto}, \citenamefont {Chiu}, \citenamefont {Chang}, \citenamefont {Wu}, \citenamefont {Perng}, \citenamefont {Chung}, \citenamefont {Kao}, \citenamefont {Yeh}, \citenamefont {Chao}, \citenamefont {Chen}, \citenamefont {Huang},\ and\ \citenamefont {Chen}}]{singh_development_2021}%
  \BibitemOpen
  \bibfield  {author} {\bibinfo {author} {\bibfnamefont {A.}~\bibnamefont {Singh}}, \bibinfo {author} {\bibfnamefont {H.~Y.}\ \bibnamefont {Huang}}, \bibinfo {author} {\bibfnamefont {Y.~Y.}\ \bibnamefont {Chu}}, \bibinfo {author} {\bibfnamefont {C.~Y.}\ \bibnamefont {Hua}}, \bibinfo {author} {\bibfnamefont {S.~W.}\ \bibnamefont {Lin}}, \bibinfo {author} {\bibfnamefont {H.~S.}\ \bibnamefont {Fung}}, \bibinfo {author} {\bibfnamefont {H.~W.}\ \bibnamefont {Shiu}}, \bibinfo {author} {\bibfnamefont {J.}~\bibnamefont {Chang}}, \bibinfo {author} {\bibfnamefont {J.~H.}\ \bibnamefont {Li}}, \bibinfo {author} {\bibfnamefont {J.}~\bibnamefont {Okamoto}}, \bibinfo {author} {\bibfnamefont {C.~C.}\ \bibnamefont {Chiu}}, \bibinfo {author} {\bibfnamefont {C.~H.}\ \bibnamefont {Chang}}, \bibinfo {author} {\bibfnamefont {W.~B.}\ \bibnamefont {Wu}}, \bibinfo {author} {\bibfnamefont {S.~Y.}\ \bibnamefont {Perng}}, \bibinfo {author} {\bibfnamefont {S.~C.}\ \bibnamefont {Chung}}, \bibinfo {author} {\bibfnamefont {K.~Y.}\
  \bibnamefont {Kao}}, \bibinfo {author} {\bibfnamefont {S.~C.}\ \bibnamefont {Yeh}}, \bibinfo {author} {\bibfnamefont {H.~Y.}\ \bibnamefont {Chao}}, \bibinfo {author} {\bibfnamefont {J.~H.}\ \bibnamefont {Chen}}, \bibinfo {author} {\bibfnamefont {D.~J.}\ \bibnamefont {Huang}}, \ and\ \bibinfo {author} {\bibfnamefont {C.~T.}\ \bibnamefont {Chen}},\ }\href {\doibase 10.1107/s1600577521002897} {\bibfield  {journal} {\bibinfo  {journal} {J. Synchrotron Radiat.}\ }\textbf {\bibinfo {volume} {28}},\ \bibinfo {pages} {977} (\bibinfo {year} {2021})}\BibitemShut {NoStop}%
\bibitem [{\citenamefont {Wang}\ \emph {et~al.}(2019)\citenamefont {Wang}, \citenamefont {Fabbris}, \citenamefont {Dean},\ and\ \citenamefont {Kotliar}}]{wang_edrixs_2019}%
  \BibitemOpen
  \bibfield  {author} {\bibinfo {author} {\bibfnamefont {Y.}~\bibnamefont {Wang}}, \bibinfo {author} {\bibfnamefont {G.}~\bibnamefont {Fabbris}}, \bibinfo {author} {\bibfnamefont {M.}~\bibnamefont {Dean}}, \ and\ \bibinfo {author} {\bibfnamefont {G.}~\bibnamefont {Kotliar}},\ }\href {\doibase 10.1016/j.cpc.2019.04.018} {\bibfield  {journal} {\bibinfo  {journal} {Comput. Phys. Commun.}\ }\textbf {\bibinfo {volume} {243}},\ \bibinfo {pages} {151} (\bibinfo {year} {2019})}\BibitemShut {NoStop}%
\bibitem [{\citenamefont {Toth}\ and\ \citenamefont {Lake}(2015)}]{toth_linear_2015}%
  \BibitemOpen
  \bibfield  {author} {\bibinfo {author} {\bibfnamefont {S.}~\bibnamefont {Toth}}\ and\ \bibinfo {author} {\bibfnamefont {B.}~\bibnamefont {Lake}},\ }\href {\doibase 10.1088/0953-8984/27/16/166002} {\bibfield  {journal} {\bibinfo  {journal} {J. Phys. Condens. Matter.}\ }\textbf {\bibinfo {volume} {27}},\ \bibinfo {pages} {166002} (\bibinfo {year} {2015})}\BibitemShut {NoStop}%
\end{thebibliography}
\end{document}